\documentclass[useAMS,usenatbib]{mn2e}
\usepackage{txfonts}
\usepackage{natbib}
\usepackage{aas_macros}
\usepackage[all]{xy}
\usepackage{multicol}

\usepackage{graphicx}

\usepackage{hyperref}
\usepackage[usenames,dvipsnames]{color}
\definecolor{darkblue}{rgb}{0,0,.5}
\hypersetup{bookmarksopen,
			pdfstartview={FitH},
			colorlinks=true,
			breaklinks=true,
			linkcolor=darkblue,
			menucolor=darkblue,
			urlcolor=Red,
			citecolor=Green,
			linkcolor=blue}
\usepackage[all]{hypcap}

\def\del#1{{}}

\newcommand{\dd}{\mathrm{d}}
\newcommand{\eqref}[1]{(\ref{#1})}

\newcommand{\kvec}{\bmath k}

\title[Intrinsic ellipticity bispectra]{A theoretical estimate of intrinsic ellipticity bispectra induced by angular momenta alignments}
\author[Philipp M. Merkel and Bj{\"o}rn Malte Sch\"afer]
{Philipp M. Merkel$^1$\thanks{e-mail: philipp.merkel@urz.uni-heidelberg.de} and Bj{\"o}rn Malte Sch\"afer$^2$\\
${}^1$Institut f{\"u}r Theoretische Astrophysik, Zentrum f{\"u}r Astronomie, Universit{\"a}t Heidelberg, Philosophenweg 12, 69120 Heidelberg, Germany\\
${}^2$Astronomisches Recheninstitut, Zentrum f{\"u}r Astronomie, Universit{\"a}t Heidelberg, Philosophenweg 12, 69120 Heidelberg, Germany}

\begin{document}

\pagerange{\pageref{firstpage}--\pageref{lastpage}}
\pubyear{2014}
\maketitle
\label{firstpage}

\begin{abstract}
Intrinsically aligned galaxy shapes are one of the most important systematics in cosmic shear measurements. So far theoretical studies of intrinsic alignments almost exclusively focus on their statistics at the two-point level. Results from numerical simulations, however, suggest that third-order measures might be even stronger affected. We therefore investigate the (angular) bispectrum of intrinsic alignments. In our fully analytical study we describe intrinsic galaxy ellipticities by a physical alignment model, which makes use of tidal torque theory. 
We derive expressions for the various combinations of intrinsic and gravitationally induced ellipticities, i.e. \textit{III}-, \textit{GII}- and \textit{GGI}-alignments, and compare our results to the shear bispectrum, the \textit{GGG}-term. The latter is computed using hyper-extended perturbation theory. Considering equilateral and squeezed configurations we find that for a \textit{Euclid}-like survey intrinsic alignments (\textit{III}-alignments) start to dominate on angular scales smaller than $20\, \arcmin$ and $13\, \arcmin$, respectively.
This sensitivity to the configuration-space geometry may allow to exploit the cosmological information contained in both the intrinsic and gravitationally induced ellipticity field.
On smallest scales ($\ell \sim 3000$) \textit{III}-alignments exceed the lensing signal by at least one order of magnitude. The amplitude of the \textit{GGI}-alignments is the weakest. It stays below that of the shear field on all angular scales irrespective of the wave-vector configuration.
\end{abstract}


\section{Introduction}
\label{sec_introduction}

The unprecedented precision being in reach of forthcoming galaxy surveys like \textit{DES}\footnote{http://www.darkenergysurvey.org/} and \textit{Euclid}\footnote{http://www.euclid-ec.org/}
brought systematic effects in weak lensing observations into focus. Key among these is the intrinsic alignment of galaxy shapes mimicking the lensing signal \citep[see][for an overview]{2012MNRAS.424.1647K}. By now, intrinsic alignments have been detected not only in numerical simulations \citep{2006MNRAS.371..750H,2007ApJ...671.1135K} but also in a number of Sloan Digital Sky Survey samples \citep{2009ApJ...694..214O, 2011MNRAS.410..844M}. Other, non-astrophysical, systematics result from the shape measurement process \citep[e.g.][]{2012MNRAS.423.3163K}  and photometric redshift errors \citep[e.g.][]{2008MNRAS.387..969A}. 

Conventional weak lensing analyses concentrate on the two-point statistics of the cosmic shear field: its correlation function or equivalently its power spectrum \citep[e.g.][]{1998MNRAS.301.1064K}. But also higher-order statistics, specifically the three-point correlation function and the corresponding shear bispectrum, contain valuable cosmological information, which is not directly accessible from measurements of the power spectrum alone. In particular, the combination of both, shear power spectrum and bispectrum, allows to break several parameter degeneracies \citep{2003MNRAS.340..580T,2010APh....32..340V}. Furthermore, the weak lensing bispectrum can be used to tighten constraints on the dark energy parameters considerably \citep{2004MNRAS.348..897T, 2013arXiv1306.4684K}. \citep[For a detailed overview of the complementary information content of the weak lensing power spectrum and bispectrum see][and references therein.]{2013MNRAS.429..344K} In addition to that, \citet{2013MNRAS.434..148S} pointed out that the combination of second- and third-order weak lensing statistics could help to discriminate between different models of the baryonic feedback from galaxy formation on the matter distribution. The most recent detection of cosmic shear statistics beyond the power spectrum has been carried out for the COSMOS survey using the third-order moment of the aperture mass measure \citep{2011MNRAS.410..143S}. 

In order to control the intrinsic alignment contamination at the two-point level a variety of methods have been developed in the past \citep{2002A&A...396..411K,2003A&A...398...23K,2003MNRAS.339..711H,2008A&A...488..829J, 2009A&A...507..105J,2010A&A...523A...1J,2013MNRAS.432.2433H}. In contrast to this, the efforts regarding higher order statistics are relatively limited (despite their aforementioned predictive power). \citet{2010A&A...523A..60S} exploited the different redshift dependence of intrinsic and lensing induced galaxy ellipticities for removing the intrinsic alignment signal. This so-called nulling technique, however, involves a nonnegligible loss of signal. Suppression of intrinsic alignment contributions while conserving the constraining power of the lensing signal is provided by the self-calibration method proposed by \citet{2012MNRAS.423.1663T}. Most recently, \citet{2014A&A...561A..68S} showed that an $E$/$B$-mode separation in third-order aperture mass statistics may serve as a viable tool in the search for weak lensing systematics like intrinsic alignments. The removal or at least the suppression of intrinsic alignment contributions to higher-order statistics appears to be of particular interest as results from numerical simulations suggest that they may exceed the lensing signal by up to an order of magnitude depending on the survey's median redshift \citep{2008MNRAS.388..991S}. Apart from its role as a contaminant in the context of gravitational lensing, the intrinsic ellipticity field carries valuable information about both galaxy  formation and evolution. Thus, a pronounced signal at the three-point level could possibly offer a way to access this information. In both cases, a profound theoretical understanding of intrinsically aligned galaxies beyond the two-point function is desirable. In this work we study analytically the bispectrum of intrinsic alignments. Our starting point is a physically motivated model for the intrinsic ellipticity field \citep{2001MNRAS.320L...7C}. This so-called quadratic alignment model has already been successfully applied in several studies on intrinsic alignments \citep{2001ApJ...559..552C, 2002MNRAS.332..788M,2004PhRvD..70f3526H,2013MNRAS.435..194C}. 

As a first step we extend the commonly used observables of the intrinsic ellipticity field, its $E$- and $B$-mode, by the scalar ellipticity, i.e. its modulus (Section~\ref{sec_intrinsic_galaxy_ellipticities_and_weak_lensing}). We then derive expressions for the angular bispectra of the various combinations of these three observables (Section~\ref{sec_ellipticity_bispectra}). In addition to the pure intrinsic ellipticity bispectra, \textit{III}-alignments (Section~\ref{subsec:intrinsic_ellipticity_bispectra}), we investigate mixed bispectra, too. These involve the $E$-mode of the cosmic shear field once and twice in case of \textit{GII}- and \textit{GGI}-alignments, respectively (Section~\ref{subsec:mixed_bispectra}). The decisive quantity our findings need ultimately to be compared to is the weak lensing bispectrum. To this end we evaluate our expressions numerically for two distinct wave-vector configurations: We consider equilateral and flattened triangles assuming a galaxy survey comparable to the forthcoming \textit{Euclid} mission (Section~\ref{subsec_results}). We conclude in Section~\ref{sec_summary}.   

Throughout this work we choose a spatially flat $w\mathrm{CDM}$ cosmology as reference. The relevant equations of the homogeneous background and (linear) structure growth are briefly summarized in Section~\ref{sec_cosmology}. To be specific we adopt for the dark energy equation-of-state parameter the value $w=-0.9$. The share of matter in the Universe's energy density amounts to~$\Omega_\mathrm{m}=0.25$ including the small contributions from the baryons~$\Omega_\mathrm{b}=0.04$. The initial fluctuations in the cold dark matter (CDM) component are assumed to be adiabatic and are characterized by the spectral index~$n_\mathrm{s}=1$ and an amplitude corresponding to~$\sigma_8=0.8$. Finally, we set for the Hubble function evaluated today~$H_0=100\, h \, \mathrm{km}\,\mathrm{Mpc}^{-1} \, \mathrm{s}^{-1}$ with $h=0.72$.

\section{Cosmology}
\label{sec_cosmology}

The expansion of the homogeneous background in a spatially flat Friedmann-Lema\^itre-Robertson-Walker universe is governed by the Hubble function 
\begin{equation}
 H^2(a) = H^2_0 \left( \frac{\Omega_{\mathrm m}}{a^{3}}  + \frac{1 - \Omega_{\mathrm m}}{a^{3(1+w)}} \right),
\end{equation}
where $\Omega_{\mathrm m}$ describes the matter content (in units of the critical density) and the equation-of-state parameter of the dark energy fluid is assumed to be constant. As time parameter we have chosen the scale factor $a$. Its relation to comoving distance $\chi$ is given by
\begin{equation}
 \chi = c\int_a^1\frac{\dd a}{a^2 H(a)}.
\end{equation}
Consequently, the Hubble distance $\chi_H=c/H_0$, setting the scale up to which Newtonian gravity is applicable, is the natural unit of (comoving) distances.

In the linear regime the evolution of the cosmic density field~$\delta$ is independent of scale and completely described by the growth function $D_+(a)$, i.e. $\delta(\kvec, a) = D_+(a)\delta_0(\kvec)$ (normalized to unity today). The growth function in turn is the solution of the growth equation \citep{1997PhRvD..56.4439T,1998ApJ...508..483W,2003MNRAS.346..573L}
\begin{equation}
 \frac{\dd^2}{\dd a^2}D_+(a) + \frac{1}{a}\left( 3 + \frac{\dd \log H}{\dd \log a}\right) \frac{\dd}{\dd a} D_+(a) = \frac{3}{2a^2}\Omega_{\mathrm m}(a) D_+(a).
\end{equation}
Being a statistically homogeneous and isotropic Gaussian random field the fluctuations of the linearly evolving density field are fully characterized by its power spectrum
\begin{equation}
 \left\langle \delta(\bmath k ) \delta^*(\bmath k')\right\rangle = (2\upi)^3 \delta_\mathrm{D}(\bmath k - \bmath k')P_{\delta\delta}(k).
\end{equation}
In this work we will be more often concerned with the statistics of the Newtonian gravitational potential $\Phi$ instead. We shall therefore make use of the (comoving) Poisson equation
\begin{equation}
 \label{eq_Poisson_equation}
 -k^2 \Phi(\bmath k,a) = \frac{3}{2}\frac{\Omega_{\mathrm m} H_0^2}{a}\delta(\bmath k,a)
\end{equation}
to mediate between both dynamical fields.
The ansatz for the matter power spectrum is a power law modulated by an appropriate transfer function
\begin{equation}
 P_{\delta\delta}(k) \propto k^{n_\mathrm{s}} T^2(k)
\end{equation}
with
\begin{eqnarray}
 \nonumber
 T(q) &=& \frac{\log (1+2.34 q)}{2.34 q} \left(1 + 3.89 q + (16.1q)^2 + (5.46q)^3 \right.\\
 && \left.+\, (6.71q)^4\right)^{-\frac{1}{4}}
\end{eqnarray}
\citep{1986ApJ...304...15B}.
Rescaling of the wave-number $q=k/\Gamma$ by the shape parameter
\begin{equation}
 \Gamma = \Omega_{\mathrm m} h \exp\left[ -\Omega_{\mathrm b}\left(1 + \frac{\sqrt{2h}}{\Omega_{\mathrm m}}   \right) \right]
\end{equation}
accounts for the influence of a nonvanishing baryon density $\Omega_{\mathrm{b}}$ \citep{1995ApJS..100..281S}.
Finally, the power spectrum is normalized to the variance of the linearly evolved density field smoothed by a top hat filter on the scale $R=8\,\mathrm{Mpc}\,h^{-1}$
\begin{equation}
 \sigma^2_R = \frac{1}{2\upi^2} \int_0^\infty k^2 \dd k\, W^2(kR) P_{\delta\delta}(k).
\end{equation}
The Fourier transform of the top hat filter can be expressed by the first order spherical Bessel function~$j_1(x)$ \citep{1972hmfw.book.....A}: 
$W(x)=3j_1(x)/x$.
Nonlinear structure growth enhances the fluctuations on small scales. The resulting corrections to the matter power spectrum are well captured by the fit suggested by \citet{2003MNRAS.341.1311S} which is gauged to the results of cosmic structure formation simulations.

\section{Intrinsic galaxy ellipticities and weak gravitational lensing}
\label{sec_intrinsic_galaxy_ellipticities_and_weak_lensing}

\subsection{Weak gravitational lensing}

Light emitted by distant galaxies is deflected by the gravitational potentials of the intervening large-scale structure. The observer detects light-rays originally starting from position $\bbeta$ at the lensed position $\btheta$.
Consequently, gravitational lensing alters the observed shape of the galaxies. In a locally linearized form this effect is captured by the Jacobian of the lens mapping
\begin{equation}
 \mathbfss A = \frac{\partial \bbeta}{\partial \btheta}=
 \left(
 \begin{array}{cc}
 1- \kappa -\gamma_+ & -\gamma_\times\\
 -\gamma_\times & 1- \kappa +\gamma_+
 \end{array}
 \right).
\end{equation}
The convergence $\kappa$ describes the isotropic change in size of the image, while the two shear components $\gamma_+$ and $\gamma_\times$ 
encode the deformation of the source galaxy. Choosing the $z$-axis as line-of-sight, the $\gamma_+$ component describes the stretching in the $x - y$ 
directions. The stretching along axes rotated by $45^\circ$ is given by $\gamma_\times$. Rotations of the image may not be generated in this linearized 
treatment
\citep[see][for a comprehensive review on gravitational lensing]{2001PhR...340..291B}.

All quantities can be computed from second derivatives of the lensing potential $\phi$ (which we formulate here in Fourier space for convenience)
\begin{equation}
 \kappa(\bmath k) = -\frac{1}{2} (k_x^2 + k_y^2)\phi(\bmath k),
 \label{eq:kappa_relation_to_lensing_potential}
\end{equation}
\begin{equation}
 \gamma_+(\bmath k) = -\frac{1}{2} (k_x^2 - k_y^2) \phi (\bmath k),
 \label{eq:gamma_plus_relation_to_lensing_potential}
\end{equation}
\begin{equation}
 \gamma_\times (\bmath k) = - k_x k_y \phi(\bmath k ).
 \label{eq:gamma_cross_relation_to_lensing_potential}
\end{equation}
The lensing potential itself is given by the line-of-sight projection of the Newtonian gravitational potential $\Phi$. In a spatially flat universe where 
(comoving) angular diameter distance and comoving distance coincide we have
\begin{equation}
 \phi (\bmath k,\chi) = 2 \int_0^\chi \chi' \dd \chi' \frac{\chi - \chi' }{\chi} \Phi(\bmath k, \chi').
\end{equation}

So far, we have not taken the redshift distribution of the lensed population of background sources into account. We therefore introduce the following weighting
\begin{equation}
 \kappa (\bmath k) = -(k_x^2 + k_y^2) \int_0^{\chi_H} \dd \chi W_{\kappa}(\chi) \Phi(\bmath k, \chi),
 \label{eq:convergence_field}
\end{equation}
where the lensing efficiency function is defined by
\begin{equation}
 W_\kappa (\chi) =\chi\int_\chi^{\chi_H} \dd \chi' n(z) \frac{\dd z}{\dd \chi'} \frac{\chi' - \chi}{\chi'}.
 \label{eq:weak_lensing_weighting_function}
\end{equation}
The expressions for the shear components are generalized in completely the same  way.
For the redshift distribution of lensed background galaxies we use the common parametrization
\begin{equation}
 n(z) = n_0 \left(\frac{z}{z_0}\right)^2 \exp \left[-\left( \frac{z}{z_0}\right)^\beta \right] \dd z
 \quad
 \mathrm{with}
 \quad
 \frac{1}{n_0} = \frac{z_0}{\beta}\Gamma\left(\frac{3}{\beta} \right).
 \label{eq:galaxy_distribution}
\end{equation}
In our analysis we choose $\beta = 3/2$ and $z_0=0.64$ corresponding to a median redshift of 0.9 as anticipated for \textit{Euclid} 
\citep{2013LRR....16....6A}. 

While the convergence is a scalar quantity, the shear components constitute a spin-2 field which is most conveniently recast in a complex notation $\gamma = \gamma_+ + \mathrm{i} \gamma_\times = |\gamma| \: \mathrm{e}^{2\mathrm{i}\varphi}$. As for any spin-2 field a decomposition in its parity conserving ($E$-mode) and parity violating ($B$-mode) part proves advantageous \citep{1996astro.ph..9149S,1997PhRvD..55.7368K,2002ApJ...568...20C}
\begin{equation}
 k^2 E(\bmath k ) = (k_x^2 - k_y^2) \gamma_+(\bmath k) + 2 k_xk_y \gamma_\times(\bmath k),
 \label{eq:curl_gradient_decomposition_E_mode}
\end{equation}
\begin{equation}
 k^2 B(\bmath k) = - 2 k_x k_y \gamma_+(\bmath k) + (k_x^2 - k_y^2) \gamma_\times(\bmath k).
 \label{eq:curl_gradient_decomposition_B_mode}
\end{equation}
Plugging in explicitly the expressions for the shear components (equation~\ref{eq:gamma_plus_relation_to_lensing_potential} and 
\ref{eq:gamma_cross_relation_to_lensing_potential}), it turns out that the $E$-mode coincides with the 
convergence~\eqref{eq:kappa_relation_to_lensing_potential}, while the $B$-mode 
identically vanishes. As a consequence, the statistics of cosmic lensing is entirely described by the (angular) power spectrum of the convergence, 
provided that the Newtonian potential can be treated as Gaussian random field. 
Carrying out an appropriate Limber projection \citep{1953ApJ...117..134L} the convergence spectrum is given by
\begin{equation}
 C^{\kappa\kappa}_\ell = \int_0^{\chi_H} \frac{\dd \chi}{\chi^6} \, W_\kappa^2(\chi) \, \ell^4P_{\Phi\Phi}(k = \ell/\chi,\chi).
\end{equation}
Statistics beyond the power spectrum only become important when the late time nonlinear growth of density perturbations is considered
(see Section~\ref{subsec:cosmic_shear_bispectrum}).

\subsection{Intrinsic galaxy ellipticities}

As seen before, cosmic lensing introduces a shear in the shape of the source galaxy. In addition to this there is also an intrinsic shear, i.e. ellipticity, of the lensed galaxy which is analogously described by a spin-2 field.
For spiral galaxies with a thin disk the intrinsic ellipticity can be related to the direction of its angular momentum $\bmath{\hat{L}} \equiv \bmath L / L$ assuming that the disc forms perpendicular to the spin axis \citep{2001ApJ...559..552C,2002MNRAS.332..788M}
\begin{equation}
 \epsilon = \epsilon_+ + \mathrm{i} \epsilon_\times = |\epsilon| \: \mathrm{e}^{2\mathrm i \varphi}
\end{equation}
with
\begin{equation}
 \epsilon_+ = \alpha\frac{\hat L^2_x - \hat L^2_y}{1+ \hat L_z^2}, \quad
 \epsilon_\times = 2\alpha \frac{\hat L_x \hat L_y}{1 + \hat L_z^2} \quad
 \mathrm{and} \quad
 |\epsilon| = \alpha \frac{\hat L^2_x + \hat L^2_y}{1+ \hat L_z^2}.
 \label{eq_intrinsic_ellipticities}
\end{equation}
Again we assume that the $z$-axis of the coordinate system coincides with the line-of-sight.
The factor $0<\alpha\leq1$ is a phenomenological measure for the relative galaxy thickness. For perfectly thin discs $\alpha = 1$. Typically, one sets $\alpha \simeq 0.75$ \citep{2001ApJ...559..552C}.
The mechanism responsible for the ellipticities of ellipticals is different: Since their total angular momentum is rather small the intrinsic ellipticity is mainly determined by the velocity dispersion along the principal axes of the three-dimensional ellipsoid. Our analysis will focus on spirals only. They are the dominating type, in particular at high redshifts, and outside clusters, i.e. in the field.

For further progress we assume now that the angular momentum of  a galaxy follows largely that of its host dark matter halo. Then the theory of tidal torques allows to relate the angular momentum acquired by the halo to the surrounding gravitational potential. Thus, both intrinsic ellipticities as well as gravitational shear are traced back to the same dynamical field. The statistics of the latter is well understood in linear theory where it is considered as Gaussian random field. In this way, it is possible to address higher order statistics of intrinsic ellipticities as well as mixed statistics involving intrinsic and extrinsic shear. But before we proceed we should mention that the key assumption of almost perfect alignment of the angular momentum of the galaxy and that of its host halo is challenged by a number of structure formation simulations \citep{2002ApJ...576...21V,2004ApJ...613L..41N,2005ApJ...627L..17B,2005ApJ...627..647B,2008ASL.....1....7M,2011arXiv1106.0538K}. 
Furthermore, \citet{2013ApJ...766L..15L} emphasized the importance of vortical flows in addition to shear flows during the advanced stages of the halo's angular momentum acquisition and \citet{2014MNRAS.440L..46A} pointed at  extensions to tidal torque theory necessary for an explanation of the hierarchical spin alignment in the cosmic web. 
Nonetheless, our ansatz is well suited for an analytical treatment which primarily aims at the understanding and possible detection of the statistical properties of a large cosmological ensemble. Our results provide upper limits on the intrinsic alignment contamination of weak lensing data.

\subsubsection{Angular momenta in tidal torque theory}

In the framework of tidal torque theory \citep{1949MNRAS.109..365H,1970Afz.....6..581D,1984ApJ...286...38W} the angular momentum of a dark matter 
halo is built up by its inertia tensor 
\begin{equation}
 I_{ij} = 
 \Omega_{\mathrm{m}} \rho_{\mathrm{crit}} a^3\int_\Gamma\dd^3 q\, (\bmath q - \bmath{\bar{q}})_i (\bmath q - \bmath{\bar{q}})_j
\end{equation}
and the tidal field tensor $\Phi_{ij} = \partial_i\partial_j\Phi$:
\begin{equation}
 \label{eq_angular_momentum}
 L_i = a^3H(a)\frac{\dd D_+}{\dd a}\varepsilon_{ijk} I_{jl} \Phi_{lk}
\end{equation}
\citep[see][for a review on galactic angular momenta]{2009IJMPD..18..173S}. 
More precisely, the angular momentum results from the misalignment of the eigenframes of the two tensors.
The inertia tensor is derived from the second moments of the mass distribution of the protohalo filling the Lagrangian volume $\Gamma$ with center of mass $\bmath{\bar{q}}$. Obviously, $\bmath q$ is a Lagrangian coordinate. Here and in the following we adopt Einstein's convention for the summation over repeated indices.
It is interesting to note that the time evolution of the angular momentum completely factorizes due to the fact that tidal torque theory makes use of the Zel'dovich approximation (\citealp{1970A&A.....5...84Z,1996MNRAS.282..436C}; see \citealp{1996MNRAS.282..455C} for an inclusion of leading-order  corrections from Lagrangian perturbation theory). Linearly evolving fields are appropriate because the formation of protogalaxies takes place at early stages. 
As an immediate consequence the direction of the angular momentum becomes time independent and thus the intrinsic galaxy ellipticity, too.

A complete analysis of the statistics, such as correlation functions, of the galaxy angular momentum described by 
equation~\eqref{eq_angular_momentum} is quite involved, even in the case of Gaussian random fields \citep{2012MNRAS.421.2751S}. Particularly, the calculation of the inertia tensor is a sophisticated task because it needs to be evaluated at a peak region in the cosmic density field from which the halo forms by gravitational collapse. In order to overcome these difficulties \citet{2002MNRAS.332..788M} proposed several simplifications the applicability of which has been verified using the results from numerical studies. Following the earlier work of \citet{2001MNRAS.320L...7C}, they assume that for any individual galaxy the eigenframe moment of inertia is the same, whereas the eigenframes follow an isotropic distribution. Supposing that two principle moments coincide the third one defines a symmetry axis $\bmath{\hat{ n}}$. Then the tensor of inertia takes the remarkably simple form 
$I_{ij} \sim \hat{n}_i\hat{n}_j$. 
Finally, \citet{2002MNRAS.332..788M} neglect any correlation between the tidal field and the tensor of inertia. This simplification is motivated by the fact that both fields exhibit different correlation lengths. Correlations in the inertia tensor mainly arise from smaller scales while those of the tidal field are long-ranged. This separation of scales allows for a successive averaging-process. First, one can average over the different possible orientations of the inertia tensor and subsequently over realizations of the tidal field. The expectation value of the angular momentum for a given tidal field then reads
\begin{equation}
 \label{eq_mean_angular_momentum}
 \langle L_i  L_j \rangle = \frac{1}{15}\left(\varepsilon_{i k l} \varepsilon_{j m n} \Phi_{ lm}\Phi_{k n} - \Phi_{i k}\Phi_{j k} + \delta_{ij}\Phi_{k l}\Phi_{k l} \right).
\end{equation}
Now, all statistical quantities can be derived from the primordial gravitational potential.

\subsubsection{Intrinsic ellipticity correlations}

In principle, it would now be possible to compute the correlation functions of the intrinsic ellipticities by combining equations~\eqref{eq_intrinsic_ellipticities} and \eqref{eq_mean_angular_momentum}. However, in order to facilitate the actual computation \citet{2002MNRAS.332..788M} suggested to drop the dependence of the intrinsic ellipticity on the $z$-component of the angular momentum and to use
\begin{equation}
 \epsilon_+ = C\left(L_x^2 - L_ y^2 \right)
 \quad
 \mathrm{and}
 \quad
 \epsilon_\times = 2C L_x L_y,
 \label{eq:intrinsic_ellipticity_field}
\end{equation}
with an appropriately chosen constant $C$, instead. \citet{2013MNRAS.435..194C} compared this ansatz to the one proposed by \citet{2001ApJ...559..552C} who used the full relations of equation~\eqref{eq_intrinsic_ellipticities} but a different angular momentum model. 
The resulting angular power spectra of the intrinsic ellipticities do not differ substantially: The functional shape of the spectra is much the same for both models whereas its amplitude (normalized to the power on largest scales, which are expected to be the least sensitive to the actual model) is slightly smaller in case of the ansatz of \citet{2002MNRAS.332..788M}. Interestingly, both models predict similar ratios of $E$- and $B$-modes. 

In our analysis we adopt a smaller numerical value for~$C$ than that used in \citet{2013MNRAS.435..194C}. It differs by a factor of five. By construction this more conservative choice corresponds to smaller correlations in the directions of the galactic angular momenta but to larger correlations between the galaxies' inertia and the tidal field of the ambient matter distribution \citep[see][for a detailed discussion]{2001ApJ...559..552C}. 

The curl/gradient decomposition of the intrinsic ellipticity field~\eqref{eq:intrinsic_ellipticity_field} according to 
equation~\eqref{eq:curl_gradient_decomposition_E_mode} and \eqref{eq:curl_gradient_decomposition_B_mode} has already been carried out by 
\citet{2002MNRAS.332..788M}:
\begin{equation}
 k^2 X(\bmath k) =  \frac{C}{15}  \int\frac{\dd^3 k'}{(2\upi)^3} f_X (\bmath k_\bot', \bmath k_\bot - \bmath k_\bot', k_z')
 \Phi_\mathcal{S}(\bmath k')\Phi_\mathcal{S}(\bmath k - \bmath k'),
 \label{eq:general_formulation_of_fields}
\end{equation}
where $X\in\lbrace E,B \rbrace$. The mode-coupling function $f_X$ contains all the information about the various derivatives of the potential and the
corresponding orientation of the wave-vectors. The explicit expression for $f_X$ is relegated to Appendix~\ref{sec_appendix_mode_coupling_functions}. Gradients along the 
line-of-sight, i.e. in $z$-direction, are neglected. Thus, $\bmath k_\bot=(k_x,k_y)$ denotes a two-dimensional wave-vector perpendicular to the 
line-of-sight. As a last technicality we briefly mention that the subscript of the Newtonian potential indicates smoothing on the scale of galaxy sized 
fluctuations (cf. equation~\ref{eq:Gaussian_smoothing}).

In addition to the $E$- and $B$-mode we now introduce the scalar ellipticity $S$
\begin{eqnarray}
 \nonumber
 S  \equiv |\epsilon| = \sqrt{\epsilon_+^2 + \epsilon_\times^2} = C \left( L_x^2 + L_y^2 \right) \hspace{3.45cm}&&\\
 =  \frac{C}{15} \biggl[
 \left(\Phi_{xx} - \Phi_{zz} \right)^2 + \left (\Phi_{yy} - \Phi_{zz} \right)^2 + 2\Phi_{xy}^2 + 5 \Phi_{xz}^2 + 5 \Phi^2_{yz} \biggr]. &&
\end{eqnarray}
The Fourier representation of the scalar mode is given by
\begin{equation}
 k^2S(\bmath k) = \frac{C}{15}  \int\frac{\dd^3 k'}{(2\upi)^3} f_S (\bmath k_\bot', \bmath k_\bot - \bmath k_\bot', k_z')
 \Phi_\mathcal{S}(\bmath k')\Phi_\mathcal{S} (\bmath k - \bmath k').
\end{equation}
in complete analogy to equation~\eqref{eq:general_formulation_of_fields}. The additional factor of $k^2$ appears just to match the corresponding 
expressions for the $E$- and $B$-mode. The explicit form of $f_S$ is given in equation~\eqref{eq_f_S}. Like the gradient mode the scalar ellipticity is a true 
scalar quantity in contrast to the curl mode which is a pseudo scalar.

In order to characterize the statistics of the scalar ellipticity we calculate its power spectrum.
To this end we invoke Wick's theorem to express the trispectrum of the gravitational potential in terms of its power spectrum. Exploiting the symmetries of 
$f_S$ given in equation~\eqref{eq_symmetries_of_f} we find the (three-dimensional) power spectrum
\begin{eqnarray}
 \nonumber
 P_{SS}(k) &=& \frac{2C^2}{225}\int\frac{\dd^3 k'}{(2\upi)^3} P_{\Phi_\mathcal{S}\Phi_\mathcal{S}} (k') 
 			P_{\Phi_\mathcal{S}\Phi_\mathcal{S}} (|\bmath k - \bmath k'|)\\
 \label{eq_power_spectrum_intrinsic_ellipticity}
 &&\times \, f^2_S(\bmath k_\bot', \bmath k_\bot - \bmath k_\bot', k_z').
\end{eqnarray}
Analogous expressions hold for the power spectra of the $E$- and $B$-mode replacing $f_S$ by $f_E$ and $f_B$, respectively \citep{2002MNRAS.332..788M}. The corresponding angular power spectrum follows as in case of cosmic shear from an appropriate Limber projection
\begin{equation}
 C^{SS} _\ell  = \int_0^{\chi_H} \frac{\dd \chi}{\chi^2} W_\epsilon^2(\chi) P_{SS}(k = \ell/\chi,\chi)
\end{equation}
but with a different weighting function
\begin{equation}
 W_{\epsilon}(\chi) = n(z) \frac{\dd z}{\dd\chi}.
\end{equation}
The galaxy distribution~$n(z)\dd z$ is that of equation~\eqref{eq:galaxy_distribution}.

There are several important differences with respect to the spectra of cosmic lensing. First of all, the gradient mode~$E$ and the scalar ellipticity~$S$ 
are not identical. Furthermore, intrinsic alignments, derived from the quadratic model, do posses a non-vanishing $B$-mode in contrast to the weak
lensing field.
Consequently, there exist cross correlations between these two, namely~$C^{SE}_\ell$. It can be obtained by replacing $f^2_S$ by 
the product $f_Sf_E$ in equation~\eqref{eq_power_spectrum_intrinsic_ellipticity}. The corresponding correlations involving the $B$-mode vanish 
identically  because these are combinations of fields with different parity. Due to the fact that in our approach intrinsic alignments are quadratic in the 
gravitational potential there is no cross correlation with the lensing induced ellipticity. The latter is linear in the Newtonian potential and thus, the 
correlation of intrinsic ellipticities and cosmic shear involves the bispectrum of the Newtonian potential, which vanishes as long as nonlinear corrections 
are discarded. Thus, spectra of the form $C^{\kappa X}_\ell$, so-called \textit{GI}-alignments, (where $X\, \in \lbrace S,E,B\rbrace$ as before) are identically zero. Correlations of this form, 
however, do arise in the so-called linear alignment model which relates the galaxy ellipticity directly to the shear tensor 
\citep{2001MNRAS.320L...7C,2004PhRvD..70f3526H}. While the quadratic ellipticity model used in this work applies primarily to spiral galaxies, the linear 
model is well suited for the description of elliptical galaxies.

\begin{figure}
 \label{fig_power_spectra}
 \centering
 \resizebox{\hsize}{!}{\includegraphics[]{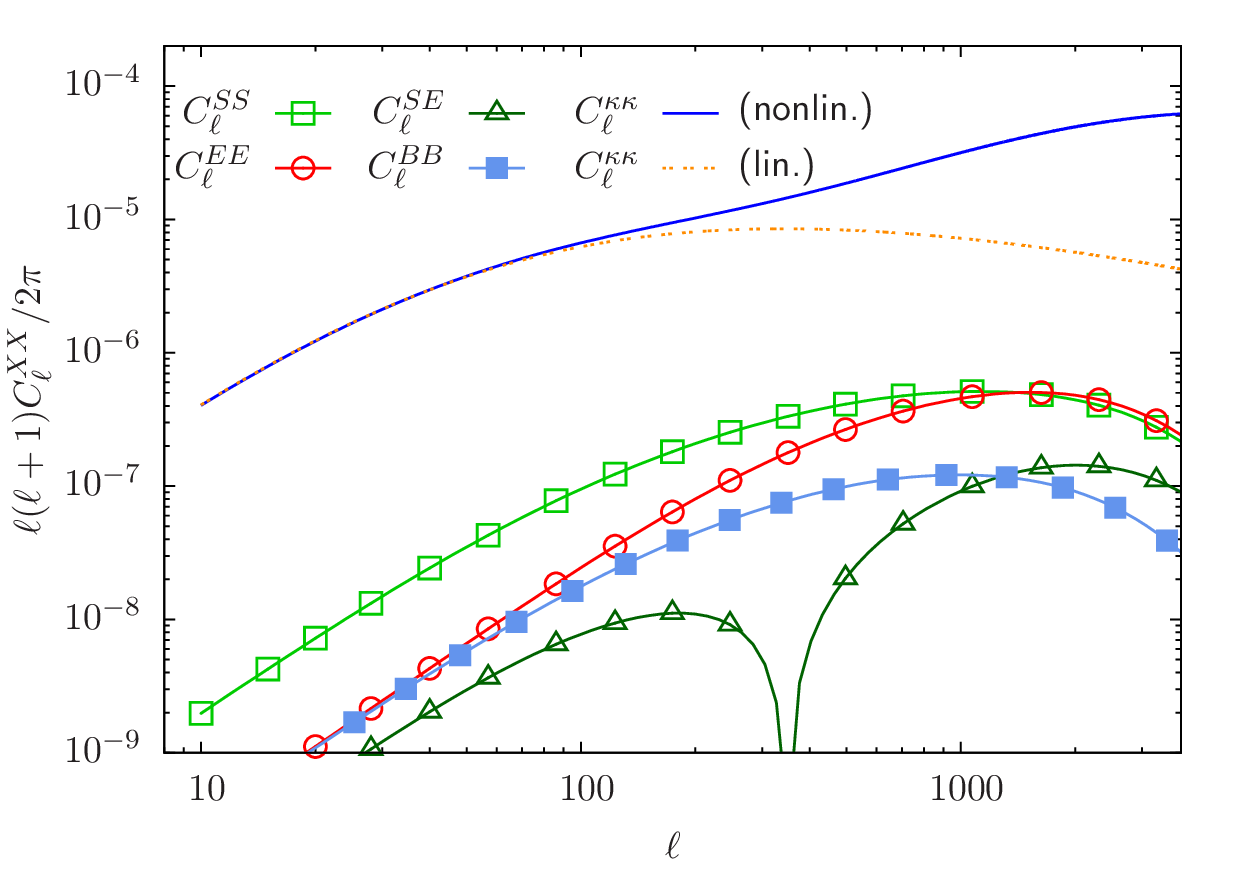}}
 \caption{Angular power spectra of the intrinsic galaxy ellipticities and the lensing induced shear. The cross spectrum of the intrinsic $S$- and $E$-mode becomes negative for $\ell \gtrsim 400$. Therefore its modulus is shown here.}
\end{figure}

In Figure~\ref{fig_power_spectra} we plot the various intrinsic ellipticity angular power spectra along with the corresponding weak lensing spectrum. The 
latter is shown for the linear and nonlinear case. For the computation of the ellipticity power spectra we applied a Gaussian filter function to the 
modes of the Newtonian potential
\begin{equation}
 \label{eq:Gaussian_smoothing}
 \Phi_\mathcal{S} (\bmath k)= \mathcal{S}_R(\bmath k) \Phi(\bmath k), \quad \mathcal{S}_R(\bmath k) =  \exp\left[-\frac{1}{2}(kR)^2\right].
\end{equation}
The smoothing scale $R$ is chosen such that galaxy-like objects of mass $M=10^{11}\, \mathrm{M}_\odot$ are selected. Hence we set 
$M=4\upi/3 \Omega_{\mathrm m} \rho_{\mathrm{crit}}R^3$.

We concentrate our discussion on the spectra of the scalar ellipticity and of its cross-correlation with the $E$-mode, which are presented for the first time
in this work. For a detailed discussion of the curl and gradient mode spectra we refer to \citet{2002MNRAS.332..788M,2013MNRAS.435..194C}.
We only note in passing that the \textit{II}-alignment signal stays below that of cosmic shear on any scale even without accounting for the additional small-scale power due to nonlinear structure growth. This is in contrast to \citet{2013MNRAS.435..194C} and stems from the fact that we employ a smaller value for the constant~$C$ in equation~\eqref{eq:intrinsic_ellipticity_field}, which enters the corresponding power spectra quadratically.

It turns out that the scalar ellipticity~$S$ is the dominant intrinsic alignment signal on all but the smallest scales $(\ell \gtrsim 800)$, where it is finally
surpassed by the gradient mode. On large scales $(\ell \lesssim 100)$ its amplitude exceeds that of the $E$-mode by about one order of magnitude.
Due to the complex mode coupling (in $k$-space) it is rather difficult to identify the different contributions to the spectra from individual multipole ranges. The different large-$\ell$ behaviour of $S$- and $E$-mode, however, may be explained as follows: The scalar ellipticity is a measure for the total (intrinsic) ellipticity of the galaxy, i.e. its deviation from a purely circular shape, whereas the gradient mode contains in addition information about the galaxy's orientation. It is physically intuitive that detailed information about the orientation is more confined to the galaxy's neighbourhood than information about the (intrinsic) shape distortion. Correlations in~$S$ are therefore longer-ranged than those in~$E$. Thus, there is less power in the $E$-mode on larger scales, i.e for smaller multipoles. This small-$\ell$ dominance of the scalar mode is also recovered for the corresponding bispcetra (see Section~\ref{subsec_results}).
Support for our interpretation can be found by considering some typical numbers for the different correlation lengths involved. The correlation scale of the intrinsic $E$-mode is about one~$\mathrm{Mpc} \, h^{-1}$ \citep{2001ApJ...559..552C,2012MNRAS.421.2751S}, while that for galaxy sized fluctuations in the cosmic large scale structure is typically five times larger \citep{2003MNRAS.346...78H}. This, however, serves only as very rough and certainly oversimplified estimate to illustrate the differences in the $S$- and $E$-mode.

The signal of the cross spectrum of scalar ellipticity~$S$ and~$E$-mode is the smallest. Only for large multipoles $(\ell \gtrsim 1000)$ it starts 
dominating over the $B$-mode. Most remarkably, the cross spectrum becomes negative for multipoles $\ell\gtrsim 400$. Thus, for angular separations
smaller than half a degree the scalar ellipticity and the gradient mode of the \textit{II}-alignments are actually anti-correlated. This underlines the different information content conveyed by~$S$ and~$E$, mentioned before. There is an important difference between general shape distortions and their orientation.

\section{Ellipticity bispectra}
\label{sec_ellipticity_bispectra}

Having computed the different power spectra for cosmic shear and the various intrinsic ellipticity fields in the last section we now address higher order 
statistics, namely the bispectrum. It is defined by
\begin{equation}
 \left\langle X(\bmath k) Y(\bmath k') Z(\bmath k'') \right\rangle = (2\upi)^3 \delta_\mathrm{D} (\bmath k + \bmath k' + \bmath k'') 
 			\: \mathcal B_{XYZ} (\bmath k, \bmath k', \bmath k'').
 \label{eq:bispectrum_fundamental_definition}
\end{equation}
Considering intrinsic ellipticities $X$, $Y$ and $Z$ can take on the values $S$, $E$ and $B$, whereas in case of weak gravitational lensing there is only 
one observable, namely  the gradient mode~$E$. In order to distinguish it from the intrinsic ellipticity $E$-mode we denote it from now on by~$\kappa$,
which is justified by the fact mentioned before that for cosmic shear the gradient mode is identically to the convergence.

\subsection{Cosmic shear bispectrum}
\label{subsec:cosmic_shear_bispectrum}

Since the convergence is linear in the density contrast its bispectrum vanishes identically if corrections due to nonlinear structure growth, which skews the 
distribution of the density field, are not accounted for. When the density contrast approaches unity its Fourier modes cease to evolve independently.
The resulting mode coupling can be computed in Eulerian perturbation theory \citep{2002PhR...367....1B,2011PhRvD..83h3518M}. In the weakly 
nonlinear regime $(\delta\lesssim 1)$ a first order calculation may suffice. Entering the highly nonlinear regime $(\delta> 1)$ one has to resort to more 
elaborate methods (see below). In order to get the tree-level contribution to the density bispectrum one has to consider the second-order density 
perturbation
\begin{equation}
 \delta^{(2)} (\bmath k, a) = \frac{D^2_+(a)}{2}\int\frac{\dd^3 k'}{(2\pi)^3} F(\bmath k', \bmath k - \bmath k')\delta(\bmath k')\delta(\bmath k -\bmath k'),
\end{equation}
with the mode-coupling function
\begin{equation}
 F(\bmath k, \bmath k') = \frac{10}{7} + \left( \frac{k}{k'} + \frac{k'}{k} \right)\: \mu
 		+ \frac{4}{7} \: \mu^2, \quad \mu \equiv \cos\sphericalangle(\bmath k,\bmath k').
 \label{eq:mode_coupling_function_tree_level}
\end{equation}
The resulting bispectrum is then given by
\begin{equation}
 \mathcal B_{\delta\delta\delta}(\bmath k_1, \bmath k_2, \bmath k_3, a) 
 = \sum_{i,j=1,2,3\atop i\neq j} D_+^4(a) F(\bmath k_i, \bmath k_j) P_{\delta\delta}(k_i) P_{\delta\delta}(k_j).
\end{equation}
Exploiting the Poisson equation~\eqref{eq_Poisson_equation} one readily relates the density bispectrum to that of the gravitational potential
\begin{equation}
 \mathcal B_{\Phi\Phi\Phi}(\bmath k_1, \bmath k_2, \bmath k_3, a) = - \left( \frac{3}{2} \Omega_\mathrm{m} H_0^2\right)^3 \frac{1}{a^3}
 				\frac{\mathcal B_{\delta\delta\delta} (\bmath k_1, \bmath k_2, \bmath k_3,a)}{k_1^2 k_2^2 k_3^2}.
 \label{eq_gravitational_potential_bispectrum}
\end{equation}
The two-dimensional flat-sky convergence bispectrum is then obtained from equation~\eqref{eq_gravitational_potential_bispectrum} by means of an 
appropriate Limber projection \citep{1953ApJ...117..134L,2004MNRAS.348..897T}
\begin{equation}
 B_{\kappa\kappa\kappa} (\bmath\ell_1, \bmath\ell_2, \bmath\ell_3) =
  -\int \frac{\dd\chi}{\chi^4} \, W_\kappa^3 (\chi) \,
  k_1^2 k_2^2 k_3^2 \, \mathcal B_{\Phi\Phi\Phi} (\bmath k_1, \bmath k_2, \bmath k_3,\chi).
  \label{eq:Limber_projection_of_convergence_bispectrum}
\end{equation}
The weighting function~$W_\kappa$ is that of equation~\eqref{eq:weak_lensing_weighting_function}. Note that we omit contributions along the 
line-of-sight, i.e. in $z$-direction. Thus, the wave-vectors entering the bispectrum on the right-hand side of 
equation~\eqref{eq:Limber_projection_of_convergence_bispectrum} have to be understood as $\bmath k_i = (\bmath\ell_i/\chi,0)$.

Besides the nonlinear evolution of cosmic structures primordial non-Gaussianities make the matter bispectrum, and hence that of the weak lensing 
convergence, non-vanishing \citep[see e.g.][for reviews on primordial non-Gaussianity]{2010CQGra..27l4010K,2010AdAst2010E..71Y}. Favouring 
squeezed configurations, as 
is evident from equation~\eqref{eq:mode_coupling_function_tree_level}, the 
structure of the matter bispectrum due to nonlinear clustering resembles that induced by primordial non-Gaussianity of local type.
The amplitude of the latter is characterized by the so-called $f_{\mathrm{NL}}$ parameter with 
$f_\mathrm{NL}\sim\mathcal{O}(10)$ \citep{2011ApJS..192...18K,2013arXiv1303.5084P}.
This, however, is about three orders of magnitude smaller than the amplitude of the structure formation induced bispectrum 
\citep[e.g.][]{2004MNRAS.348..897T,2012MNRAS.421..797S}. Contributions from primordial non-Gaussianities are therefore not considered in the 
following.

\subsection{Intrinsic ellipticity bispectra}
\label{subsec:intrinsic_ellipticity_bispectra}

We have seen that one can form naturally three observables from the intrinsic ellipticity field, the modulus $S$ and the $E$- and
$B$-mode of the complex intrinsic ellipticity field. 
All three fields are related to the Newtonian gravitational potential via
\begin{equation}
 X(\bmath k ) = \frac{1}{15} \frac{C}{k^2} \int \frac{\dd^3 k'}{(2\upi)^3}  f_X(\bmath k_\bot', \bmath k_\bot - \bmath k_\bot', k'_z)
 					\Phi_\mathcal{S} (\bmath k ') \Phi_\mathcal{S}( \bmath k - \bmath k'),
 \label{eq:general_expression_for_S_E_and_B}
\end{equation}
for $X\in \left\lbrace S,E,B\right\rbrace$. 
Since the Newtonian potential enters quadratically in 
equation~\eqref{eq:general_expression_for_S_E_and_B}, the bispectrum contains the six-point function of the (smoothed) linear gravitational potential. 
Therefore the intrinsic ellipticity bispectrum does not vanish. As discussed before this is opposed to the three-point function of the cosmic shear field,
which is only nonzero provided that nonlinear clustering is taken into account.

The first step in the computation of the intrinsic ellipticity bispectrum is to break up the six-point function of the Newtonian potential
\begin{eqnarray}
 \nonumber
 \mathcal{F}_{\bmath k_1' \bmath k_2' \bmath k_3'}^{\bmath k_1 \bmath k_2 \bmath k_3} &\equiv&
 \left\langle  \Phi_\mathcal{S}(\bmath k_1') \Phi_\mathcal{S}(\bmath k_1 - \bmath k_1') \Phi_\mathcal{S} (\bmath k_2')
 			\Phi_\mathcal{S}(\bmath k_2 - \bmath k_2') \right.\\
	&& \times \left.\Phi_\mathcal{S} (\bmath k_3') \Phi_\mathcal{S}(\bmath k_3 - \bmath k_3') \right\rangle
\end{eqnarray}
into its reducible part by means of Wick's theorem. Neglecting terms which contribute only the zeroth mode, i.e. terms proportional to 
$\delta_\mathrm{D}(\bmath k)$, we are left with eight different terms
\begin{eqnarray}
 \nonumber
 \mathcal{F}_{\bmath k_1' \bmath k_2' \bmath k_3'}^{\bmath k_1 \bmath k_2 \bmath k_3} &=&
 				\delta^\mathrm{D}_{\bmath k_1' + \bmath k_2 - \bmath k_2' }
				\delta^\mathrm{D} _{\bmath k_3' + \bmath k_1 - \bmath k_1' }
				\delta^\mathrm{D}_{\bmath k_2' + \bmath k_3 - \bmath k_3'}
				P_{\Phi_\mathcal{S}\Phi_\mathcal{S}} ( k_1' ) 
				P_{\Phi_\mathcal{S}\Phi_\mathcal{S}} ( k_2' )\\
	\nonumber
	&& \times\,	P_{\Phi_\mathcal{S}\Phi_\mathcal{S}} ( k_3' ) + \bmath k_1,\, \bmath k_1' \leftrightarrow \bmath k_2,\, \bmath k_2' \\
	\nonumber
	&&   + \,		\delta^\mathrm{D}_{\bmath k_1' + \bmath k_2'}
	    			\delta^\mathrm{D} _{\bmath k_3' + \bmath k_1 - \bmath k_1' }
				\delta^\mathrm{D}_{\bmath k_2 - \bmath k_2' + \bmath k_3 - \bmath k_3'}
				P_{\Phi_\mathcal{S}\Phi_\mathcal{S}} ( k_1' )  
				P_{\Phi_\mathcal{S}\Phi_\mathcal{S}} ( k_3' ) \\
	&& \times\,
				P_{\Phi_\mathcal{S}\Phi_\mathcal{S}} \left(\left|\bmath k_2 -\bmath k_2' \right|\right)
				+ 5 \, \mathrm{perm.}
\end{eqnarray}
Here we have introduced the abbreviation
\begin{equation}
 \delta^\mathrm{D}_{\bmath k + \ldots + \bmath k' } \equiv (2\upi)^3 \delta_\mathrm{D} (\bmath k + \ldots + \bmath k')
\end{equation}
for notational convenience. We then may rewrite the left-hand side of equation~\eqref{eq:bispectrum_fundamental_definition} as
\begin{eqnarray}
 \nonumber
 \left\langle X(\bmath k_1) Y(\bmath k_2) Z(\bmath k_3) \right\rangle
	= \frac{1}{k_1^2 k_2^2 k_3^2} \frac{C^3}{15^3}
 \prod_{i=1}^3
 \int\frac{\dd^3 k'_i}{(2\upi)^3}
 \mathcal{F}_{\bmath k_1' \bmath k_2' \bmath k_3'}^{\bmath k_1 \bmath k_2 \bmath k_3} &&\\
 \nonumber
 \times\, f_X (\bmath k_1'^\bot, \bmath k_1^\bot - \bmath k_1'^\bot, k'^z_1) 
 		f_Y (\bmath k_2'^\bot, \bmath k_2^\bot - \bmath k_2'^\bot, k'^z_2)  &&\\
\times\, f_Z (\bmath k_3'^\bot, \bmath k_3^\bot - \bmath k_3'^\bot, k'^z_3)
		\phantom{f_Y (\bmath k_2'^\bot, \bmath k_2^\bot - \bmath k_2'^\bot, k'^z_2)} &&
\end{eqnarray}
and carry out two of the $k$-integrations. Consequently, the intrinsic ellipticity bispectrum takes the following form
\begin{equation}
 \mathcal B_{XYZ} (\bmath k_1, \bmath k_2, \bmath k_3) = \frac{1}{k_1^2 k_2^2 k_3^2} \frac{C^3}{15^3} 
 		\sum_{i=1}^8 \mathcal{Q}^{(i)}_{XYZ}(\bmath k_1, \bmath k_2, \bmath k_3),
 \label{eq:three_dimensional_intrinsic_eliipticity_bispectrum}
\end{equation}
where the functions $\mathcal{Q}^{(i)}_{XYZ}$ are schematically given by
\begin{eqnarray}
 \nonumber
 \mathcal{Q}^{(i)}_{XYZ}(\ldots) &=& \int\frac{\dd^3 k}{(2\upi)^3} f_X( \ldots ) f_Y( \ldots ) f_Z( \ldots )
 			P_{\Phi_\mathcal{S}\Phi_\mathcal{S}}(\ldots)\\
 &&\times \, P_{\Phi_\mathcal{S}\Phi_\mathcal{S}}(\ldots) P_{\Phi_\mathcal{S}\Phi_\mathcal{S}}(\ldots).
\end{eqnarray}
The full expressions can be found in Appendix~\ref{sec_appendix_source_functions}.

The two-dimensional flat-sky bispectrum is obtained from equation~\eqref{eq:three_dimensional_intrinsic_eliipticity_bispectrum} via the
corresponding Limber projection (cf. equation~\ref{eq:Limber_projection_of_convergence_bispectrum})
\begin{equation}
 B_{XYZ} (\bmath\ell_1, \bmath\ell_2, \bmath\ell_3) =
  \int \frac{\dd\chi}{\chi^4} W_\epsilon^3 (\chi) \, 
  \mathcal B_{XYZ} (\bmath\ell_1/\chi, \bmath\ell_2/\chi, \bmath\ell_3/\chi).
  \label{eq:Limber_projection_of_intrinsic_elllipticity_bispectrum}
\end{equation}

\subsection{Cosmic shear-intrinsic ellipticity bispectra}
\label{subsec:mixed_bispectra}

So far we have considered the bispectra of the intrinsic and lensing induced ellipticities separately. But mixed bispectra exist, too. One can distinguish 
two different cases: \textit{GGI}- and \textit{GII}-alignments. \textit{GGI}-alignments occur when the light of two background galaxies is distorted by a lens 
whose shear field aligns a foreground galaxy at the same time. Accordingly, \textit{GII}-alignments arise from the physical situation of two intrinsically 
aligned galaxies in the foreground and a third lensed background galaxy. The foreground galaxies need to be close-by in redshift as well as in angular separation since the typical correlation length of \textit{II}-alignments is of the order of about one $\mathrm{Mpc} \,h^{-1}$ 
\citep[e.g.][]{2001ApJ...559..552C,2012MNRAS.421.2751S}.
Configurations where the intrinsically aligned galaxy or galaxy pairs reside in the background are not expected to contribute because this would require 
a matter structure being extremely largely extended along the line-of-sight. 

For the quadratic alignment model there are only \textit{GGI}-alignments present in linear theory because \textit{GII}-alignments involve the correlator of  
an odd number of potential modes. Odd correlators, however, vanish if inhomogeneous clustering is not taken into account. Thus, the linear contributions to the mixed bispectra are given by correlators like $\left\langle\kappa(\bmath k_1) \kappa(\bmath k_2) X(\bmath k_3) \right\rangle$. Including tree-level corrections to the convergence gives rise to \textit{GII}-alignments due to terms of the form
$\langle\kappa^{(2)}(\bmath k_1) X(\bmath k_2) Y(\bmath k_3)\rangle$ ($X, Y\in\left\lbrace S,E,B\right\rbrace$). These corrections contribute also 
to the \textit{GGI}-alignments via for instance $\langle\kappa^{(2)}(\bmath k_1) \kappa^{(2)}(\bmath k_2) X(\bmath k_3)\rangle$. However, 
being quadratic in $\kappa^{(2)}$ these terms are higher-order corrections and will not be considered in the following. It is interesting to note that
different perturbation orders do not intermingle, i.e. $\langle \kappa(\bmath k_1) \kappa^{(2)}(\bmath k_2) X(\bmath k_3)\rangle = 0$ etc.

\subsubsection{\textit{GGI}-alignments}

In order to compute the \textit{GGI}-type bispectra we first bring the expression for the convergence field~\eqref{eq:convergence_field} into a form similar 
to that of equation~\eqref{eq:general_expression_for_S_E_and_B}
\begin{equation}
 \kappa(\bmath k) = f_\kappa ( \bmath k_\bot) \Phi(\bmath k)
 \label{eq:convergence_using_f_kappa}
\end{equation}
with
\begin{equation}
 f_\kappa (\bmath k_\bot  ) = -\left( k^2_x + k^2_y \right).
\end{equation}
Here we have dropped the lensing kernel for simplicity. It will be restored later, when we carry out the Limber projection to obtain the flat-sky bispectra.
Prior to that we have to evaluate the following correlator
\begin{eqnarray}
 \nonumber
 \left\langle \kappa (\bmath k_1) \kappa (\bmath k_2) X(\bmath k_3) \right\rangle =
 \frac{C}{15} \frac{f_\kappa(\bmath k^\bot_{1}) f_\kappa(\bmath k^\bot_{2})}{k_3^2}
 \int\frac{\dd^3k_4}{(2\upi)^3}\left\langle \Phi (\bmath k_1) \Phi (\bmath k_2)\right. &&\\
\times  \left.\Phi_\mathcal{S} (\bmath k_4) \Phi_{\mathcal{S}} (\bmath k_3 - \bmath k_4) \right\rangle 
 	f_X(\bmath k^\bot_4, \bmath k^\bot_3 - \bmath k^\bot_4, k_4^z). &&
\end{eqnarray}
We note that the structure is the very same as that of the tree-level matter bispectrum. Thus, after applying Wick's theorem no $k$-integration
remains:
\begin{eqnarray}
 \nonumber
  \mathcal B_{\kappa\kappa X}(\bmath k_1, \bmath k_2, \bmath k_3 ) &=&
 \frac{2}{15} \frac{C}{k_3^2} f_\kappa(\bmath k^\bot_{1}) f_\kappa(\bmath k^\bot_{2})
f_X(-\bmath k_1^\bot, -\bmath k_2^\bot, k_1^z )\\
 &&
 \times\,\mathcal{S}_R(-\bmath k_1) \mathcal{S}_R(-\bmath k_2)
 P_{\Phi\Phi}( k_1) P_{\Phi\Phi}(k_2).
\label{eq:mixed_bispectra_GGI}
\end{eqnarray}
In order to simplify the expression we have made use of equation~\eqref{eq_symmetries_of_f}.

As before the flat-sky bispectra are obtained from an appropriate Limber projection of equation~\eqref{eq:mixed_bispectra_GGI}. However, in case of the 
mixed bispectra we have to account for the different weighting functions and different time evolution of the fields under consideration
\begin{eqnarray}
 \nonumber
  B_{\kappa\kappa X}(\bmath\ell_1, \bmath\ell_2, \bmath\ell_3 ) &=& \int_{0}^{\chi_H} \frac{\dd\chi}{\chi^4}
 W^2_\kappa(\chi)  W_\epsilon(\chi) \frac{D^2_+(a)}{a^2}\\
 &&\qquad\qquad\quad\times \, \mathcal B_{\kappa\kappa X}(\bmath \ell_1 / \chi, \bmath\ell_2/\chi,\bmath \ell_3/\chi).
  \label{eq:Limber_projection_of_GGI_bispectrum}
\end{eqnarray}
At this point we should note that the equilateral \textit{GGI}-bispectrum involving the curl mode of the intrinsic ellipticity field is identically zero. This is an 
immediate consequence of the fact that 
$\mathcal B_{\kappa\kappa B} (\bmath\ell_1/\chi,\bmath\ell_2/\chi,\bmath\ell_3/\chi) \sim f_B(\bmath\ell_1/\chi, \bmath\ell_2/\chi,0) \sim 
\left(\ell^2_1 - \ell^2_2\right)$ (cf. equation~\ref{eq:mode_coupling_function_f_B}). 

\subsubsection{\textit{GII}-alignments}

Since the bispectra of \textit{GII}-type contain the tree-level convergence the corresponding expressions become much more involved. As in the case of 
the pure intrinsic ellipticity bispectra one has to evaluate the six-point function of the Newtonian potential. In order to use the results of 
Section~\ref{subsec:intrinsic_ellipticity_bispectra} we aim at adopting the functional form of equation~\eqref{eq:general_expression_for_S_E_and_B} for 
the tree-level convergence. 

We start in complete analogy to equation~Ê\eqref{eq:convergence_using_f_kappa} with 
\begin{equation}
 \kappa^{(2)}(\bmath k) = f_\kappa(\bmath k_\bot) \Phi^{(2)} (\bmath k),
\end{equation}
where the linearity of the Poisson equation~\eqref{eq_Poisson_equation} guaranties that
\begin{equation}
 \Phi^{(2)} (\bmath k) = - \frac{3}{2} \frac{\Omega_\mathrm{m} H_0^2}{k^2 a} \delta^{(2)} (\bmath k).
\end{equation}
In the following we omit the time dependence of the convergence field for clarity. It will be reestablished in the final expression for the Limber projection.
Defining the mode coupling function by
\begin{eqnarray}
 \nonumber
 f_{\kappa^{(2)}}(\bmath k, \bmath k') &=& -\left(3\Omega_\mathrm{m} H_0^2 \right)^{-1}\frac{1}{k^2}f_\kappa(\bmath k_\bot) 
 F(\bmath k', \bmath k - \bmath k' ) k'^2 \left(\bmath k - \bmath k' \right)^2\\
	&&  \times\, \mathcal{S}_R^{-1} (\bmath k') \mathcal{S}_R^{-1}(\bmath k - \bmath k')
\end{eqnarray}
we arrive at
\begin{equation}
 \kappa^{(2)}(\bmath k) = \int\frac{\dd^3 k'}{(2\upi)^3} f_{\kappa^{(2)}}(\bmath k, \bmath k')
 	\Phi_\mathcal{S}(\bmath k')\Phi_\mathcal{S}(\bmath k - \bmath k').
\end{equation}
Thus, the corresponding \textit{GII}-bispectra are given by
\begin{equation}
 \mathcal B_{\kappa^{(2)}XY}(\bmath k_1,\bmath k_2, \bmath k_3) = \frac{1}{k_1^2 k_2^2 k_3^2} \frac{C^2}{15^2} 
 			\sum_{i=1}^8 \mathcal Q^{(i)}_{\kappa^{(2)}XY}(\bmath k_1, \bmath k_2, \bmath k_3).
\end{equation}
Restoring the correct time evolution the Limber projection then reads
\begin{eqnarray}
 \nonumber
 B_{\kappa^{(2)}XY}(\bmath \ell_1,\bmath \ell_2, \bmath \ell_3) &=&
  \int\frac{\dd\chi}{\chi^4}W_\kappa(\chi) W^2_\epsilon(\chi)\frac{D_+^2(a)}{a}\\
  &&\qquad\qquad\times \, \mathcal B_{\kappa^{(2)}XY}(\bmath\ell_1/\chi,\bmath\ell_2/\chi, \bmath\ell_3/\chi).
  \label{eq:Limber_projection_of_GII_bispectrum}
\end{eqnarray}

From these findings the derivation of the next higher correction to the \textit{GGI}-alignments is straightforward. We only state the relevant expressions 
here for completeness
\begin{eqnarray}
 \nonumber
 B_{\kappa^{(2)}\kappa^{(2)}X}(\bmath \ell_1,\bmath \ell_2, \bmath \ell_3) &=&
  \int\frac{\dd\chi}{\chi^4}W^2_\kappa(\chi) W_\epsilon(\chi)\frac{D_+^4(a)}{a^2}\\
  &&\qquad\quad\times \, \mathcal B_{\kappa^{(2)}\kappa^{(2)}X}(\bmath\ell_1/\chi,\bmath\ell_2/\chi, \bmath\ell_3/\chi),
\end{eqnarray}
with
\begin{equation}
 \mathcal B_{\kappa^{(2)} \kappa^{(2)} X}(\bmath k_1,\bmath k_2, \bmath k_3) = \frac{1}{k_1^2 k_2^2 k_3^2} \frac{C}{15} 
 			\sum_{i=1}^8 \mathcal Q^{(i)}_{\kappa^{(2)} \kappa^{(2)} X}(\bmath k_1, \bmath k_2, \bmath k_3).
\end{equation}
Before we move on we would like to briefly contemplate the expressions for the various types of bispectra. Comparing eqution~\eqref{eq:Limber_projection_of_convergence_bispectrum}, \eqref{eq:Limber_projection_of_intrinsic_elllipticity_bispectrum}, \eqref{eq:Limber_projection_of_GGI_bispectrum} and \eqref{eq:Limber_projection_of_GII_bispectrum} we notice that they differ in three distinct aspects, namely in their time evolution (attributed to structure growth), their mode-coupling structure as well as in their dependency on redshift (manifest in the different weighting or efficiency functions). 

\subsection{Results}
\label{subsec_results}

\begin{figure}
 \centering
 \resizebox{\hsize}{!}{\includegraphics[]{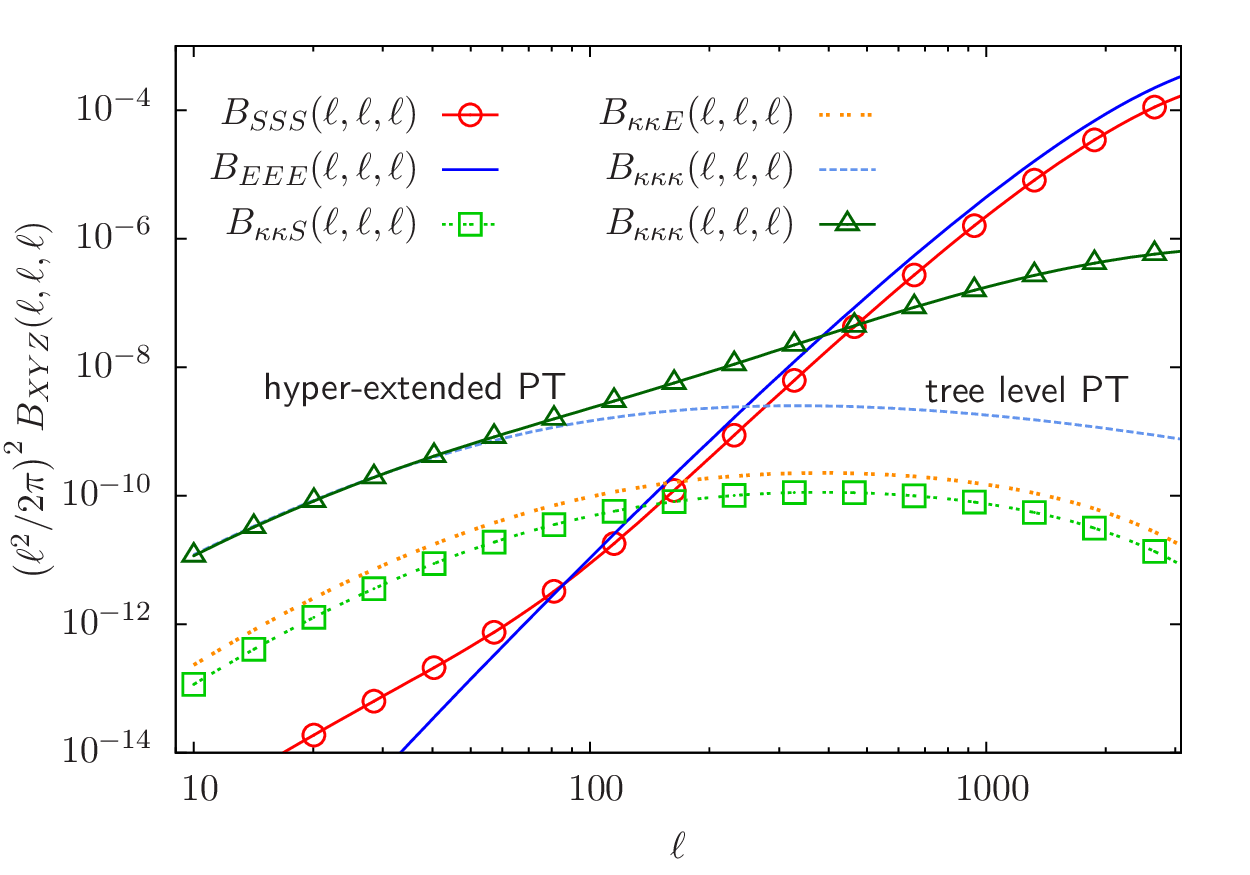}}
 \caption[Equilateral ellipticity bispectra.]{Equilateral bispectra of the gradient (solid blue) and scalar (red circles) mode of the intrinsic ellipticity field along with
 			 the weak lensing bispectrum as obtained from tree level (light blue dashed) and hyper-extended (dark green triangles) perturbation theory. 
			 In addition the non-vanishing mixed bispectra (\textit{GGI}-alignments) are shown.}
 \label{fig:ellipticity_bispectra}
\end{figure}

\begin{figure}
 \centering
 \resizebox{\hsize}{!}{\includegraphics[]{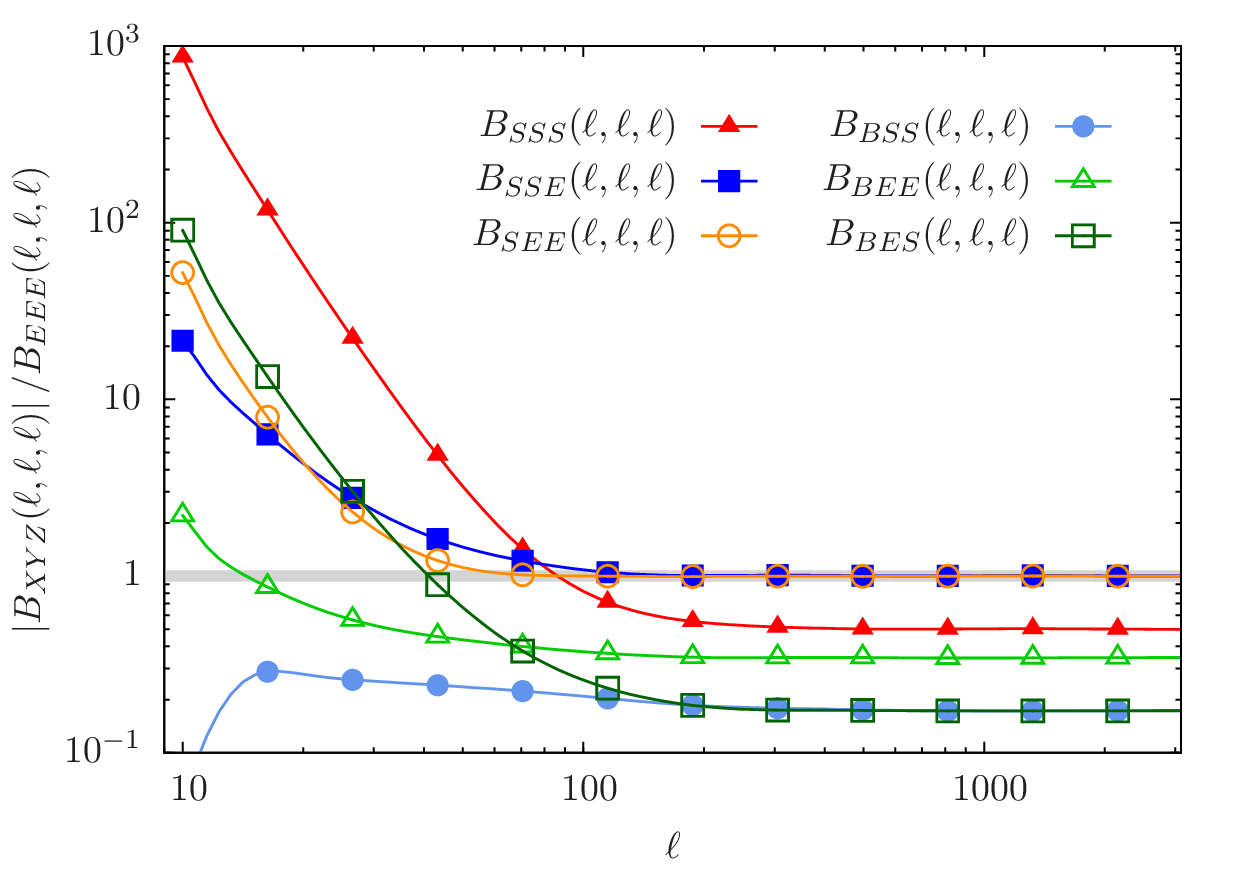}}
 \caption[Equilateral intrinsic ellipticity bispectra for various field combinations]{Equilateral intrinsic ellipticity bispectra for various field combinations. 
 					All spectra are normalized to the amplitude of the pure $E$-mode bispectrum.}
 \label{fig:intrinsic_ellipticity_bispectra_ratios}
\end{figure}

In order to illustrate our results we first focus on the equilateral configuration. We show in Figure~\ref{fig:ellipticity_bispectra} the intrinsic 
ellipticity bispectra for the scalar and the gradient mode, $S$ and $E$, respectively, as an example. 
Their functional form is typical for all bispectra which can be built from 
the various combinations of the three ellipticity fields $S$, $E$ and $B$. This is demonstrated in Figure~\ref{fig:intrinsic_ellipticity_bispectra_ratios}, 
where we plot several mixed bispectra normalized to that of the gradient mode being of largest amplitude.
Bispectra involving an odd number of curl modes do not vanish in general due to parity (in contrast to the corresponding power spectra). This can be most easily seen in the full-sky formalism \citep[e.g.][]{2000PhRvD..62d3007H}. However, we observe that bispectra containing one or more $B$-modes are significantly suppressed. The reason for this suppression can be found in the mode coupling function~$f_B$ (equation~\ref{eq:mode_coupling_function_f_B}). In contrast to the corresponding expressions for the scalar and gradient mode it is directly proportional to the difference in wave-vectors (more precisely to their modulus squared) and therefore subjected to substantial cancellations, which do not occur for the other two mode coupling functions. This suppression for curl modes has already been encountered in Figure~\ref{fig_power_spectra}, where the various ellipticity power spectra are shown. Accordingly, the more curl modes are included the stronger the suppression of the corresponding bispectrum. Bispectra with more than one $B$-mode are practically zero and are therefore not shown in Figure~\ref{fig:intrinsic_ellipticity_bispectra_ratios}.

The amplitude of the intrinsic ellipticity bispectra is tremendously large. Figure~\ref{fig:ellipticity_bispectra} suggests that it is by far the dominant 
small-scale signal even for our rather conservative choice for~$C$. It exceeds the cosmic shear signal obtained from first order perturbation theory by about four orders of magnitude for 
$\ell \sim 1000$. On these scales, however, the applicability of (tree level) perturbation theory ultimately breaks down and more elaborated methods 
need to be employed. These are hyper-extended perturbation theory \citep{1999ApJ...520...35S} on the one hand and the halo model approach 
\citep{2002PhR...372....1C} on the other hand.
The accuracy of these models reaches the 10-30 per cent level with respect to the amplitude of the three-point correlation function 
\citep{2003MNRAS.340..580T,2003MNRAS.344..857T}.
In this work we use the fitting formula of \citet{2001MNRAS.325.1312S} for the density bispectrum. This formula is based on hyper-extended perturbation theory and we supply it with the nonlinear matter power spectrum as suggested by \citet{2004MNRAS.348..897T}.
Taking the density fluctuations enhanced by nonlinear 
structure growth into account the lensing signal increases significantly. The difference to the result from tree-level perturbation theory amounts to more 
than three orders of magnitude on the smallest scales. Nonetheless the signal of the intrinsic ellipticity bispectrum is still much larger on these angular 
scales. Cosmic shear dominates only on scales larger than $20\,\arcmin$, i.e. for $\ell \lesssim 600$. 
Finally, one notices from Figure~\ref{fig:intrinsic_ellipticity_bispectra_ratios} that intrinsic ellipticity bispectra involving the scalar mode~$S$ are enhanced on large scales with respect to the pure $E$-mode spectrum (cf. our discussion at the end of Section~\ref{sec_intrinsic_galaxy_ellipticities_and_weak_lensing}), whereas those containing vortical modes are suppressed on small angular scales.

\begin{figure}
 \centering
 \resizebox{\hsize}{!}{\includegraphics[]{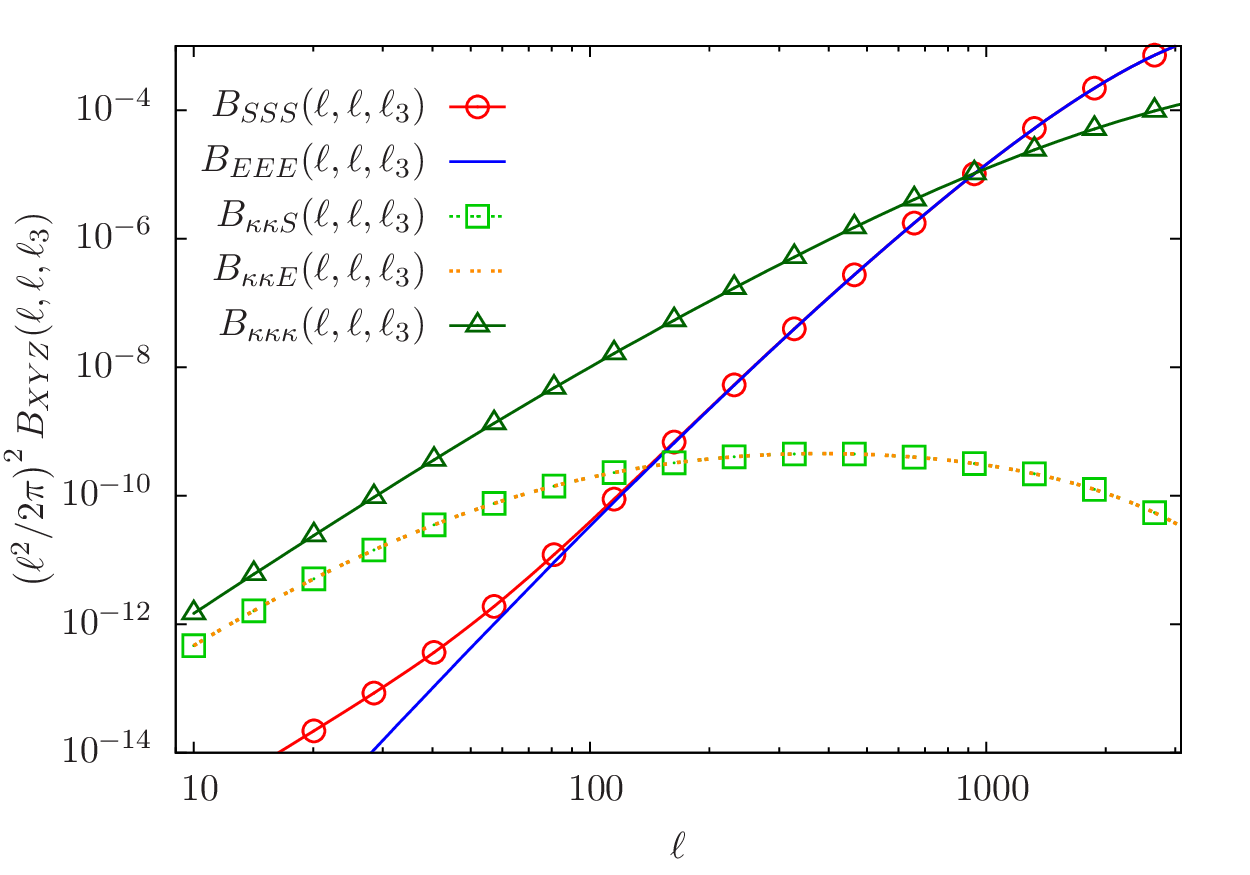}}
 \caption[Squeezed ellipticity bispectra.]{Intrinsic ellipticity and cosmic shear bispectra for a squeezed setup of the involved wave-vectors. To be specific we set~$\ell_1=\ell_2\equiv \ell$ and $\cos\sphericalangle(\bmath\ell_1,\bmath\ell_2) \simeq -1$. Note that the tree level weak lensing bispectrum has been omitted in this plot. The colour code is that of Figure~\ref{fig:ellipticity_bispectra}.}
 \label{fig:ellipticity_bispectra_squeezed_configuration}
\end{figure}

In order to investigate the geometrical dependence of the shape of the bispectra we consider squeezed configurations next. In this case two of the wave-vectors are almost perfectly anti-parallel making the third one nearly vanish. Note that in case of mixed bispectra one has to interchange the wave-vector and field index simultaneously, in other words $B_{XYZ}(\ell_1,\ell_2,\ell_3)$ is expected to be different from $B_{XZY}(\ell_1,\ell_2,\ell_3)$. To be specific we set $\ell_1 =\ell_2\equiv\ell$ and $\cos\sphericalangle(\bmath\ell_1,\bmath\ell_2) \simeq -1$ in the following but we have confirmed that the results for other representative choices are quite similar. We first focus on a comparison of \textit{III}-, \textit{GGI}- and \textit{GGG}-alignments (Figure~\ref{fig:ellipticity_bispectra_squeezed_configuration}). 
\begin{figure}
 \centering
 \resizebox{\hsize}{!}{\includegraphics[]{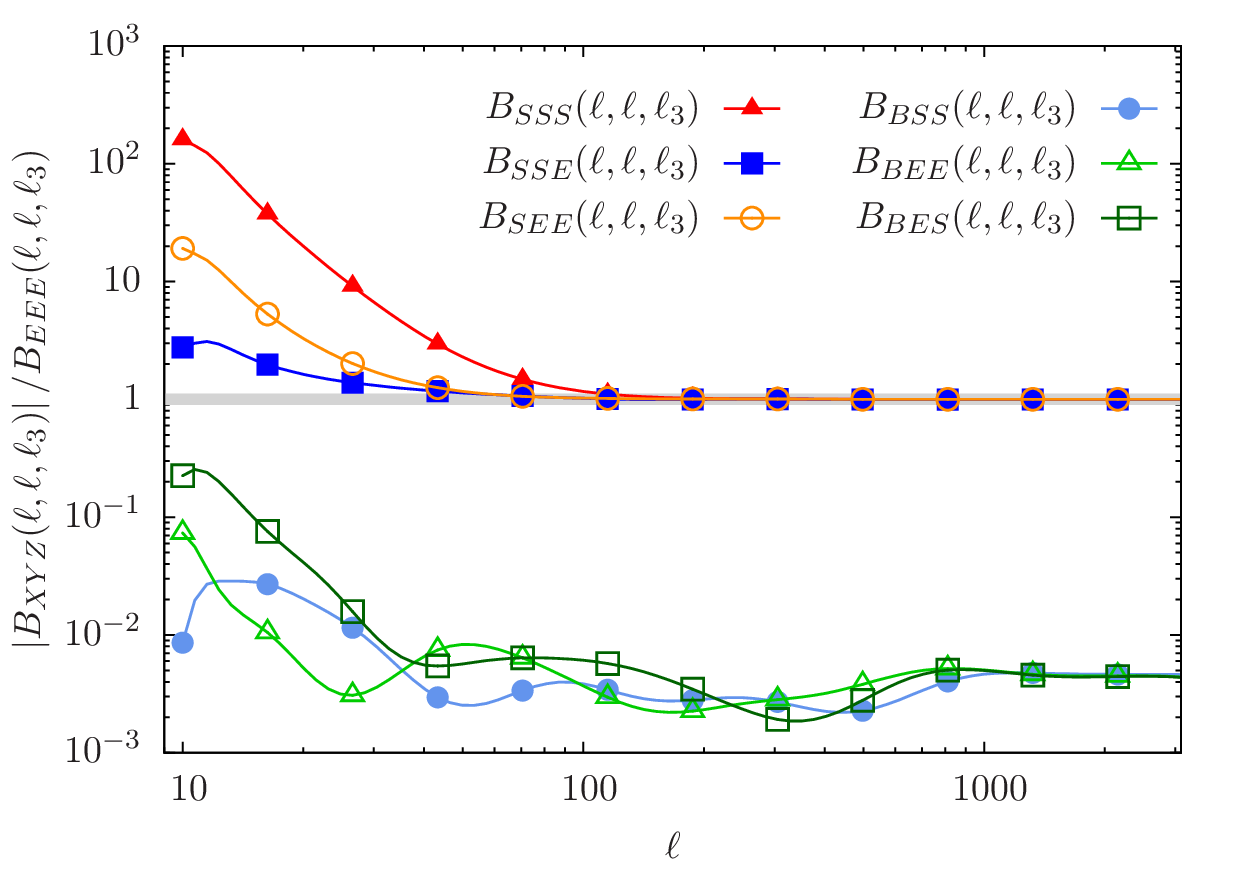}}
 \caption[Squeezed intrinsic ellipticity bispectra for various field combinations]{Squeezed intrinsic ellipticity bispectra for various field combinations. 
 					The modulus of two of the wave-vectors is~$\ell$ and the angle between these two is very close to~$\upi$. 
					All spectra are again normalized to the amplitude of the pure $E$-mode bispectrum.}
 \label{fig:intrinsic_ellipticity_bispectra_ratios_squeezed_configuration}
\end{figure}
As before we choose the auto spectra of the gradient and scalar mode as representatives of the intrinsic ellipticity field. We address \textit{III}-alignments of mixed type in Figure~\ref{fig:intrinsic_ellipticity_bispectra_ratios_squeezed_configuration}.

Looking at Figures~\ref{fig:ellipticity_bispectra_squeezed_configuration} and~\ref{fig:intrinsic_ellipticity_bispectra_ratios_squeezed_configuration} we see that the differences between the $S$- and $E$-mode bispectra are almost completely gone. Only the discrepancy on large scales $(\ell \lesssim 100)$ remains but less pronounced. Accordingly \textit{GGI}-alignments containing either scalar or gradient modes become indistinguishable for the squeezed configuration. Since the matter bispectrum favours flattened configurations (cf. equation~\ref{eq:mode_coupling_function_tree_level}) the changes in the cosmic shear bispectrum are most prominent. It is enhanced by more than two orders of magnitude. At the same time the amplitude of the \textit{III}-alignments increase only by a factor of about ten. Thus, the  huge relative difference between \textit{III}- and \textit{GGG}-alignments found for equilateral configurations is considerably attenuated. In particular, the \textit{III}-signal starts dominating on much smaller scales $(\ell\sim 900)$. Squeezed configurations, therefore, provide direct access to the small-scale cosmic shear bispectrum even in the presence of intrinsic alignments.

Figure~\ref{fig:intrinsic_ellipticity_bispectra_ratios_squeezed_configuration} shows that for squeezed configurations \textit{III}-alignments including either scalar or gradient modes are virtually identical on sub-degree scales in accordance with our previous finding for \textit{GGI}-alignments. Furthermore, we notice that the suppression of bispectra containing $B$-modes is even stronger than in case of equilateral spectra. Here they are almost three orders of magnitude smaller than the pure $E$-mode bispectrum.

Before we go on we shall comment on the amplitude of the \textit{III}-alignments. It is important to keep in mind that our results can provide no more than an estimate limited by both the assumptions having led to the model invoked for the intrinsic galaxy shapes and the vague constraints on the model parameter~$C$. This constant enters our expressions cubed and they are therefore markedly sensitive to the particular choice of its value. Thus, the amplitude of the presented ellipticity spectra is of considerable uncertainty. Nevertheless, also a more cautious choice for~$C$, i.e. a smaller value, would not change the general result that intrinsic alignments are dominant or at least comparable to the weak lensing bispectrum on small angular scales confirming results from numerical simulations \citep{2008MNRAS.388..991S}.

\begin{figure}
 \centering
 \resizebox{\hsize}{!}{\includegraphics[]{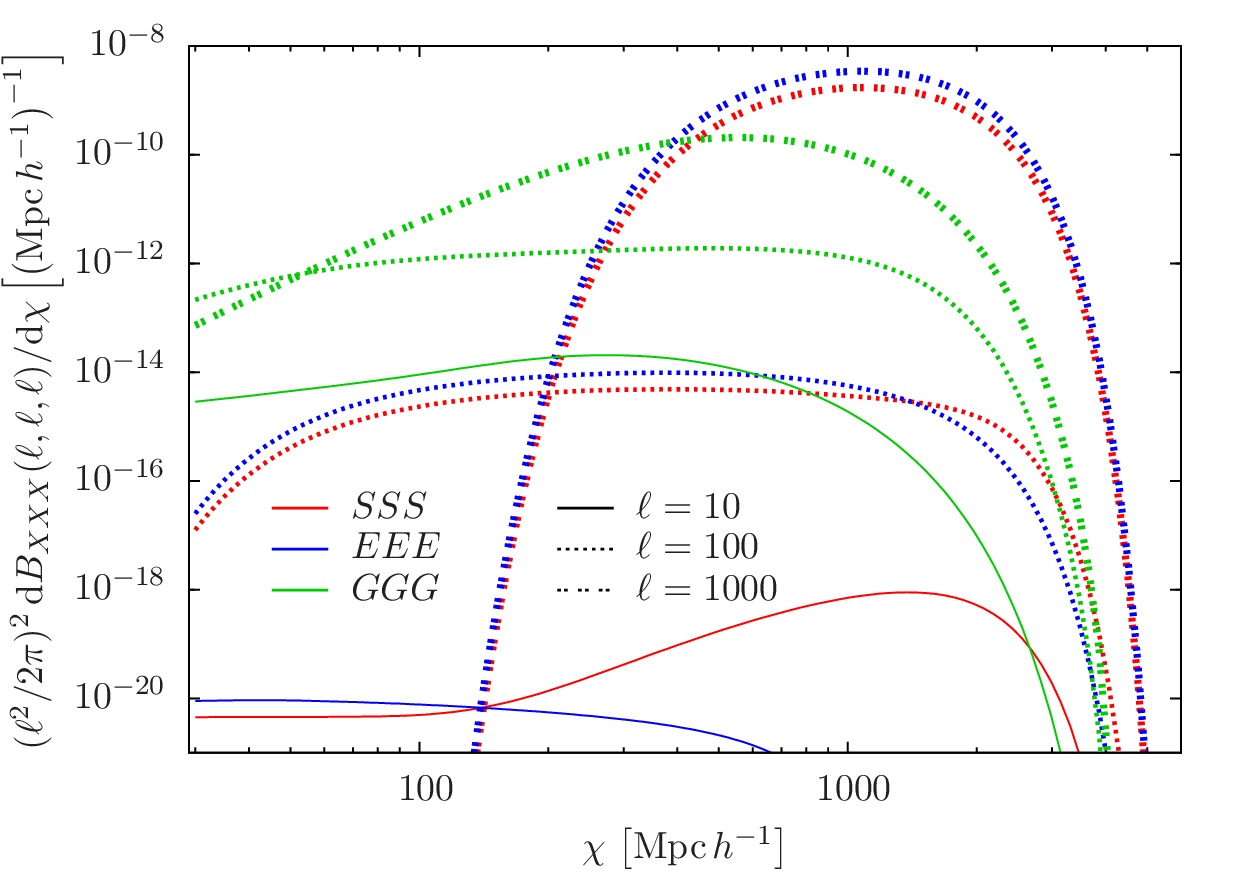}}
 \caption[Contributions to the intrinsic ellipticity and cosmic shear bispectra as a function of comoving distance]{Contributions to the intrinsic ellipticity and cosmic shear bispectra as a function of comoving distance for three different multipoles.}
 \label{fig:contributions_to_bispectra}
\end{figure}

In Figure~\ref{fig:contributions_to_bispectra} we take a look at the contribution~$\dd B_{XXX}(\ell,\ell,\ell)/\dd\chi$ to the weak lensing and intrinsic ellipticity bispectra ($S$- and $E$-mode), respectively,  as a function of comoving distance. To be specific we show the equilateral configuration for three different multipoles~($\ell = 10, \, 100, \, 1000$).
One can easily retrace how the ellipticity bispectra start to dominate for small angular scales, while being subdominant on large scales.
It is interesting to note that the comoving distance range, which contributes substantially to the \textit{III}-alignments, shrinks with increasing multipole order. This is in contrast to the contributions to the shear signal. Its functional form rather stays the same for all three multipoles under consideration. Furthermore, Figure~\ref{fig:contributions_to_bispectra} suggests that in comparison to the lensing signal the two ellipticity bispectra receive contributions also from smaller wave-numbers of the underlying cosmic density field.

\section{Summary}
\label{sec_summary}

Intrinsically aligned galaxy shapes are considered as one of the most severe contaminants in weak lensing measurements. However, most investigations of intrinsic alignments consider their statistics at the two-point level only. In this paper we extended the analysis to the three-point level and derived analytical expressions for intrinsic ellipticity bispectra induced by angular momenta alignments. 
\begin{enumerate}
\item In order to describe the intrinsic galaxy shapes we employed a physical alignment model which is quadratic in the cosmic tidal field \citep{2001MNRAS.320L...7C}. Consequently, our expressions for \textit{III}-alignments involve the six-point function of the primordial gravitational potential. Restricting ourselves to Gaussian initial conditions we made use of Wick's theorem to express the intrinsic ellipticity bispectra in terms of the primordial power spectrum. Assuming Gaussianity implies that there are no \textit{GII}-alignments to first order in our approach. \textit{GGI}-alignments, however, do exist. 
\item While the statistical properties of the cosmic shear field are fully described by the convergence or $E$-mode, respectively, there are in principle three different observables in case of intrinsic alignments. These are, in addition to the $E$-mode, the scalar ellipticity~$S$ and the $B$-mode, which is, to lowest order, identical zero for cosmic shear. 
\item Since the scalar ellipticity has been considered in this work for the first time we also presented its two-point statistics. The functional form of its power spectrum is quite similar to that of the other two observables. It is the dominant contribution to the intrinsic alignment signal on large angular scales. Due to parity only its cross spectrum with the gradient mode is different from zero. 
\item We presented analytical expressions for the bispectra of all possible combinations of the three intrinsic ellipticity field components (\textit{III}-alignments) as well as for the bispectra resulting from their combination with the weak lensing convergence (\textit{GII}- and \textit{GGI}-alignments, respectively). To illustrate our findings we employed equilateral as well as squeezed configurations and compared the results to the convergence bispectrum. The latter was computed in hyper-extended perturbation theory in order to account for the additional small-scale power due to nonlinear structure growth. For the survey specifications we chose the \textit{Euclid} mission as a reference.
\item In case of equilateral configurations we found that \textit{III}-alignments start dominating on angular scales smaller than $20\, \arcmin$, whereas their signal is negligible for small multipoles. There, i.e. for $\ell \lesssim 200$, \textit{GGI}-alignments are much more prominent but they are more than two orders of magnitude smaller than the convergence bispectrum. In general, it turned out that for all relevant scales spectra containing the gradient mode of the intrinsic ellipticity field are slightly enhanced with respect to those involving its scalar mode instead. On the smallest scales ($\ell \sim 3000$) \textit{III}-alignments exceed the cosmic shear signal by about two orders of magnitude. 
\item The situation is different for squeezed configurations. Though intrinsic alignments remain the strongest signal on smallest scales the difference with respect to \textit{GGG}-alignments reduces to a factor of ten. More important, their domination sets in on much smaller scales $(\ell \sim 900)$. This enlarges significantly the angular range where the cosmic shear bispectrum is directly accessible. In addition we find almost no differences between the bispectra made of either the scalar or gradient mode on sub-degree scales. The suppression of $B$-mode bispectra, also present in the equilateral setup, is even further enhanced for flattened configurations.
\item Because of their high amplitude intrinsic alignments are expected to be a severe contaminant in cosmic shear measurements at the three-point level. The contamination is much stronger than in case of the power spectrum. However, the distinct geometrical dependence of \textit{III}-alignments found in this work allows to mitigate their contaminating effect. While the lensing signal may be faithfully recovered from squeezed configurations on intermediate angular scales $(\ell \lesssim 500-600)$, the very strong \textit{III}-signal present in equilateral configurations may help to improve on their physical modeling. For instance it could be useful in discriminating between the so-called linear and quadratic alignment model because the bispectrum of the former is identically zero to first order. Another application one could think of is the determination of the model parameter~$C$, which is widely unconstrained by theory. Improved knowledge of this parameter would in turn alleviate the separation of intrinsic and gravitationally induced ellipticities at the two-point level. Furthermore, one might even use the additional information on \textit{III}-alignments to extend the analysis of squeezed cosmic shear bispectra towards larger multipoles.
\end{enumerate}
Finally, we would like to emphasize that our results can only serve as an estimate of the expected bispectra. Besides the simplifying assumptions necessary for an analytical description of the intrinsic galaxy shapes it is mainly the poor information about the parameter~$C$ which determines the uncertainty of our results. Since it enters the expressions for the \textit{III}-alignments to the third power the bispectra are highly sensitive to its particular value. But despite these limitations, our analytical estimates consolidate previous results from numerical simulations \citep{2008MNRAS.388..991S} promoting third-order statistics as promising way to investigate intrinsic alignments and to distinguish them from weak gravitational lensing.

\section*{Acknowledgements}

In the numerical part of our work we made extensively use of the integration routines provided by the \textsc{cuba}-library \citep{2005CoPhC.168...78H}.
PhMM acknowledges funding from the Graduate Academy Heidelberg and support from the International Max Planck
Research School for Astronomy and Cosmic Physics in Heidelberg as well as from the Heidelberg Graduate School of Fundamental Physics.
We would like to thank the anonymous referee for her/his concise comments and valuable suggestions.

\bibliography{bibtex/aamnem,bibtex/references}
\bibliographystyle{mn2e}

\appendix

\section{Mode coupling functions}
\label{sec_appendix_mode_coupling_functions}

The three different modes derived from the intrinsic galaxy ellipticity field, $S$, $E$ and $B$, all have the same structure
\begin{equation}
 X(\bmath k ) = \frac{1}{15} \frac{C}{k^2} \int \frac{\dd^3 k'}{(2\upi)^3} \Phi (\bmath k ') \Phi( \bmath k - \bmath k') f_X(\bmath k_\bot', 
 					\bmath k_\bot - \bmath k_\bot', k'_z) 
\end{equation}
for $X\in \left\lbrace S,E,B\right\rbrace$.
The corresponding coupling functions are given by
\begin{equation}
 \label{eq_f_S}
 f_S (\bmath a, \bmath b, c) = \left\lbrace c^2 \left[2c^2 - (\bmath a + \bmath b)^2 - 3(\bmath a \cdot \bmath b) \right] 
 +(\bmath a \cdot \bmath b)^2
 \right\rbrace (\bmath a + \bmath b)^2,
\end{equation}
\begin{eqnarray}
\nonumber
 f_E (\bmath a, \bmath b, c) \hspace{-0.15cm}&=& \hspace{-0.15cm}\frac{1}{2} \left(2c^2 - a^2\right) \left[b^4 + (\bmath a \cdot \bmath b)^2 - (\bmath a \times \bmath b)^2 
 			+ 2b^2 (\bmath a \cdot \bmath b)\right] \\
			\nonumber
 &&\hspace{-0.15cm}+\frac{1}{2} \left(2c^2 - b^2 \right) \left[ a^4 + (\bmath a \cdot \bmath b)^2 - (\bmath a \times \bmath b)^2  
 				+ 2a^2 (\bmath a \cdot \bmath b) \right] \\
 && \hspace{-0.15cm} 
 	+ 3c^2 \left[ (\bmath a + \bmath b)^2(\bmath a\cdot \bmath b) + 2(\bmath a \times \bmath b)^2\right]
\end{eqnarray}
and
\begin{equation}
 f_B (\bmath a, \bmath b, c) = \left(c^2 - \bmath a \cdot \bmath b \right) \left(a^2 - b^2\right)(\bmath a \times \bmath b).
 \label{eq:mode_coupling_function_f_B}
\end{equation}
Here $\bmath a$ and $\bmath b$ denote two-dimensional vectors and $c$ is a real number. Obviously, $f_S$ and $f_E$ are scalars whereas $f_B$ is a pseudo scalar. All three functions obey the following symmetries
\begin{equation}
 f_X(\bmath a, \bmath b, c) = f_X(\bmath a, \bmath b, -c) = f_X(\bmath b, \bmath a, c).
  \label{eq_symmetries_of_f}
\end{equation}

\section{Source functions}
\label{sec_appendix_source_functions}

In this appendix we gather the explicit expressions of the eight different configurations sourcing the intrinsic ellipticity bispectrum~\eqref{eq:three_dimensional_intrinsic_eliipticity_bispectrum}. They read
\begin{eqnarray}
	\nonumber
	 \mathcal Q^{(1)}_{XYZ}(\bmath k_1, \bmath k_2, \bmath k_3) =
 			\int\frac{\dd^3 k}{(2\upi)^3}
 			f_X\left( \bmath k^{\bot}, \bmath k_1^\bot - \bmath k^\bot, k^z\right)
			\hspace{1.75cm}
			&&\\
			\nonumber
 			\quad\times \, 
			f_Y\left( \bmath k^\bot + \bmath k_2^\bot, - \bmath k^\bot, k^z + k_2^z\right)
			f_Z\left( \bmath k^\bot - \bmath k_1^\bot, - \bmath k^\bot - \bmath k_2^\bot, k^z - k_1^z  \right)&&\\
			\times \,
 			P_{\Phi_\mathcal{S}\Phi_\mathcal{S}}\left( k \right)
			P_{\Phi_\mathcal{S}\Phi_\mathcal{S}}\left( \left| \bmath k + \bmath k_2 \right| \right) 
			P_{\Phi_\mathcal{S}\Phi_\mathcal{S}}\left( \left| \bmath k - \bmath k_1 \right| \right),
			\hspace{1.8cm}
			&&
\end{eqnarray}
\begin{eqnarray}
	\nonumber
	 \mathcal Q^{(2)}_{XYZ}(\bmath k_1, \bmath k_2, \bmath k_3) = 
 			\int\frac{\dd^3 k}{(2\upi)^3}
 			f_X\left( \bmath k^{\bot}, \bmath k_1^\bot - \bmath k^\bot, k^z\right)
			\hspace{1.75cm}
			&&\\
			\nonumber
 			\quad\times \,
 			f_Y\left( \bmath k^\bot - \bmath k_1^\bot, - \bmath k^\bot - \bmath k_3^\bot, k^z - k_1^z\right)
			f_Z\left( \bmath k^\bot + \bmath k_3^\bot, - \bmath k^\bot, k^z + k_3^z  \right) &&\\
			\times \,
 			P_{\Phi_\mathcal{S}\Phi_\mathcal{S}}\left( k \right)
			P_{\Phi_\mathcal{S}\Phi_\mathcal{S}}\left( \left| \bmath k + \bmath k_3 \right| \right) 
			P_{\Phi_\mathcal{S}\Phi_\mathcal{S}}\left( \left| \bmath k - \bmath k_1 \right| \right),
			\hspace{1.8cm}
			&&
\end{eqnarray}
\begin{eqnarray}
	\nonumber
	 \mathcal Q^{(3)}_{XYZ}(\bmath k_1, \bmath k_2, \bmath k_3) = 
 			\int\frac{\dd^3 k}{(2\upi)^3}
 			f_X\left( \bmath k^{\bot}, \bmath k_1^\bot - \bmath k^\bot, k^z\right)
			\hspace{1.75cm}
			&&\\
			\nonumber
 			\quad\times \,
 			f_Y\left( -\bmath k^\bot,  \bmath k_2^\bot + \bmath k^\bot, -k^z\right)
			f_Z\left( \bmath k^\bot - \bmath k_1^\bot, -\bmath k_2^\bot- \bmath k^\bot, k^z - k_1^z  \right) &&\\
			\times \,
 			P_{\Phi_\mathcal{S}\Phi_\mathcal{S}}\left( k \right)
			P_{\Phi_\mathcal{S}\Phi_\mathcal{S}}\left( \left| \bmath k - \bmath k_1 \right| \right)
			P_{\Phi_\mathcal{S}\Phi_\mathcal{S}}\left( \left| \bmath k + \bmath k_2 \right| \right),
			\hspace{1.4cm}
			&&
\end{eqnarray}
\begin{eqnarray}
	\nonumber
	 \mathcal Q^{(4)}_{XYZ}(\bmath k_1, \bmath k_2, \bmath k_3) = 
 			\int\frac{\dd^3 k}{(2\upi)^3}
 			f_X\left( \bmath k^{\bot}, \bmath k_1^\bot - \bmath k^\bot, k^z\right)
			\hspace{1.75cm}
			&&\\
			\nonumber
 			\quad\times \,
 			f_Y\left( \bmath k^\bot - \bmath k_1^\bot,  -\bmath k^\bot - \bmath k_3^\bot, k^z-k^z_1\right)
			f_Z\left( -\bmath k^\bot, \bmath k^\bot + \bmath k_3^\bot, -k^z  \right) &&\\
			\times \,
 			P_{\Phi_\mathcal{S}\Phi_\mathcal{S}}\left( k \right)
			P_{\Phi_\mathcal{S}\Phi_\mathcal{S}}\left( \left| \bmath k - \bmath k_1 \right| \right)
			P_{\Phi_\mathcal{S}\Phi_\mathcal{S}}\left( \left| \bmath k + \bmath k_3 \right| \right),
			\hspace{1.4cm}
			&&
\end{eqnarray}
\begin{eqnarray}
	\nonumber
	 \mathcal Q^{(5)}_{XYZ}(\bmath k_1, \bmath k_2, \bmath k_3) = 
 			\int\frac{\dd^3 k}{(2\upi)^3}
 			f_X\left( \bmath k^{\bot}, \bmath k_1^\bot - \bmath k^\bot, k^z\right)
			\hspace{1.75cm}
			&&\\
			\nonumber
 			\quad\times \,
 			f_Y\left( -\bmath k^\bot,  \bmath k^\bot + \bmath k_2^\bot, -k^z\right)
			f_Z\left( -\bmath k_2^\bot - \bmath k^\bot, \bmath k^\bot - \bmath k_1^\bot, -k_2^z-k^z  \right) &&\\
			\times \,
 			P_{\Phi_\mathcal{S}\Phi_\mathcal{S}}\left( k \right)
			P_{\Phi_\mathcal{S}\Phi_\mathcal{S}}\left( \left| \bmath k - \bmath k _1\right| \right)
			P_{\Phi_\mathcal{S}\Phi_\mathcal{S}}\left( \left| \bmath k + \bmath k_2 \right| \right),
			\hspace{1.6cm}
			&&
\end{eqnarray}
\begin{eqnarray}
	\nonumber
	 \mathcal Q^{(6)}_{XYZ}(\bmath k_1, \bmath k_2, \bmath k_3) = 
 			\int\frac{\dd^3 k}{(2\upi)^3}
 			f_X\left( \bmath k^{\bot}, \bmath k_1^\bot - \bmath k^\bot, k^z\right)
			\hspace{1.75cm}
			&&\\
			\nonumber
 			\quad\times \,
 			f_Y\left( -\bmath k^\bot - \bmath k_3^\bot,  \bmath k^\bot - \bmath k_1^\bot, k^z + k_3^z\right)
			f_Z\left( \bmath k^\bot  + \bmath k_3^\bot,  -\bmath k^\bot, k^z+k_3^z  \right) &&\\
			\times \,
 			P_{\Phi_\mathcal{S}\Phi_\mathcal{S}}\left( k \right)
			P_{\Phi_\mathcal{S}\Phi_\mathcal{S}}\left( \left| \bmath k - \bmath k _1\right| \right)
			P_{\Phi_\mathcal{S}\Phi_\mathcal{S}}\left( \left| \bmath k + \bmath k_3 \right| \right),
			\hspace{1.8cm}
			&&
\end{eqnarray}
\begin{eqnarray}
	\nonumber
	 \mathcal Q^{(7)}_{XYZ}(\bmath k_1, \bmath k_2, \bmath k_3) = 
 			\int\frac{\dd^3 k}{(2\upi)^3}
 			f_X\left( \bmath k^{\bot}, \bmath k_1^\bot - \bmath k^\bot, k^z\right)
			\hspace{1.75cm}
			&&\\
			\nonumber
 			\quad\times \,
 			f_Y\left( -\bmath k^\bot - \bmath k_3^\bot,  \bmath k^\bot - \bmath k_1^\bot, k^z + k_3^z\right)
			f_Z\left( -\bmath k^\bot,  \bmath k^\bot + \bmath k_3^\bot, k^z  \right) &&\\
			\times \,
 			P_{\Phi_\mathcal{S}\Phi_\mathcal{S}}\left( k \right)
			P_{\Phi_\mathcal{S}\Phi_\mathcal{S}}\left( \left| \bmath k - \bmath k _1\right| \right)
			P_{\Phi_\mathcal{S}\Phi_\mathcal{S}}\left( \left| \bmath k + \bmath k_3 \right| \right)
			\hspace{1.3cm}
			&&
\end{eqnarray}
and
\begin{eqnarray}
	\nonumber
	 \mathcal Q^{(8)}_{XYZ}(\bmath k_1, \bmath k_2, \bmath k_3) = 
 			\int\frac{\dd^3 k}{(2\upi)^3}
 			f_X\left( \bmath k^{\bot}, \bmath k_1^\bot - \bmath k^\bot, k^z\right)
			\hspace{1.75cm}
			&&\\
			\nonumber
 			\quad\times \,
 			f_Y\left( \bmath k^\bot + \bmath k_2^\bot,  -\bmath k^\bot, k^z + k_2^z\right)
			f_Z\left( -\bmath k^\bot - \bmath k_2^\bot,  \bmath k^\bot - \bmath k_1^\bot,  k^z  +k_2^z\right) &&\\
			\times \,
 			P_{\Phi_\mathcal{S}\Phi_\mathcal{S}}\left( k \right)
			P_{\Phi_\mathcal{S}\Phi_\mathcal{S}}\left( k \right)
			P_{\Phi_\mathcal{S}\Phi_\mathcal{S}}\left( \left| \bmath k + \bmath k_2 \right| \right).
			\hspace{2.6cm}
			&&
\end{eqnarray}

\bsp
\label{lastpage}

\end{document}